\documentclass[aps,nofootinbib,superscriptaddress]{revtex4}
\usepackage{amstext,amsmath,amssymb,amsfonts,bbm}
\usepackage[latin1]{inputenc}
\usepackage{fancyhdr}
\usepackage[dvips]{graphicx}
\usepackage{epsfig}
\usepackage{hyperref}

\def\bra{\langle} \def\ket{\rangle}
\def\f{\frac}
\newcommand{\SU}{\mathrm{SU}}
\newcommand{\SO}{\mathrm{SO}}

\newcommand{\U}{\mathrm{U}}
\newcommand{\lalg}[1]{\mathfrak{#1}}
\newcommand{\su}{\lalg{su}} 
 
\newcommand{\so}{\lalg{so}} 
 \newcommand{\spin}{\lalg{spin}}

\def\tl{\widetilde}
\def\pp{\partial}
 
\def\eps{\epsilon}

\newcommand{\tr}{\mathrm{tr}}
\newcommand{\Tr}{\mathrm{Tr}}

\def\C{{\mathbbm C}}

\def\be{\begin{equation}}
\def\ee{\end{equation}}
\def\bes{\begin{eqnarray}}
\def\ees{\end{eqnarray}}

\def\l({\left(}
\def\r){\right)}
\def\lv{\lvert}
\def\rv{\rvert}

\begin{document}

\title{{\large A Lagrangian approach to the Barrett-Crane spin foam model}}

\author{Valentin Bonzom}\email{valentin.bonzom@ens-lyon.fr}
\affiliation{Centre de Physique Th\'eorique, CNRS-UMR 6207,  Luminy Case 907, 13007 Marseille, France EU}
\affiliation{Laboratoire de Physique, ENS Lyon, CNRS-UMR 5672, 46 All\'ee d'Italie, 69007 Lyon, France EU}

\author{Etera R. Livine}\email{etera.livine@ens-lyon.fr}
\affiliation{Laboratoire de Physique, ENS Lyon, CNRS-UMR 5672, 46 All\'ee d'Italie, 69007 Lyon, France EU}

\begin{abstract}
We provide the Barrett-Crane spin foam model for quantum gravity with a discrete action principle,
consisting in the usual BF term with  discretized simplicity constraints which in the continuum
turn topological BF theory into gravity. The setting is the same as usually considered in the
literature: space-time is cut into 4-simplices, the connection describes how to glue these
4-simplices together and the action is a sum of terms depending on the holonomies around each
triangle. We impose the discretized simplicity constraints on disjoint tetrahedra and we show how
the Lagrange multipliers distort the parallel transport and the correlations between neighbouring simplices. We then construct the discretized BF action using a non-commutative $\star$-product
between $\SU(2)$ plane waves. We show how this naturally leads to the Barrett-Crane model. This
clears up the geometrical meaning of the model. We discuss the natural generalization of this
action principle and the spin foam models it leads to. We show how the recently introduced spinfoam
fusion coefficients emerge with a non-trivial measure. In particular, we recover the
Engle-Pereira-Rovelli spinfoam model by weakening the discretized simplicity constraints. Finally,
we identify the two sectors of Plebanski's theory and we give the analog of the Barrett-Crane model
in the non-geometric sector.

\end{abstract}
\maketitle

\section*{Introduction}

Striving to make sense of a quantum theory of gravitation, several approaches have converged
towards the construction of the spin foam formalism as a promising non-perturbative framework to
describe quantum theories in a background-independent context. A spin foam is ultimately a
combinatorial object given by a 2-complex, possibly seen as dual to a triangulation, whose
simplices are labeled by data from the representation theory of a given group. If loop quantum
gravity has given a clear picture of the kinematics of the theory in terms of spin networks, spin
foams are expected to provide a representation of their (gauge) dynamics, generated by the
Hamiltonian constraint, and thus to give a tool to define transition amplitudes \cite{SF LQG}. On
the other hand, it has been shown that path integrals for lattice gauge theories can be formulated
as sums over spin foams living on the lattice \cite{SF path int}. In particular, spin foams living
on a single triangulation are well adapted to topological BF theories \cite{baez, ooguri}. In the
3d case, where $\SU(2)$ BF theory corresponds to gravity, they have been shown to yield the correct
quantization \cite{alej karim}. The group representation data defining each spin foam can be
interpreted, using the canonical analysis, as quantized flux observables \cite{PRTVO}.

Several models have applied for describing quantum gravity. Among them, the most studied is the
Barrett-Crane (BC) model \cite{BC}, the first non-topological model. The study of its
semi-classical limit has revealed that the two-point function exhibits the correct behaviour with
regards to the distance \cite{semi class}, but that it fails to reproduce the expected tensorial
structure of the free graviton propagator of spin 2 \cite{BC fail}. New propositions have been made
to correct the drawbacks of the BC model \cite{EPR, consistently solving, FKLS, alexandrov},
including a Lorentzian version \cite{epr lorentz}, and with a finite Immirzi parameter \cite{epr
imm}. The edification of these works is based on different foundations. First, there exists a
reformulation of the action of general relativity as a topological BF theory with additional
constraints $\mathcal{C}$ quadratic in the 2-form $B\in\so(4)$:
\be \label{plebanski}
S_{GR}(B,A) = S_{BF}(B,A) + \lambda\ \mathcal{C}(B)
\ee
where $\lambda$ stands for the Lagrange multipliers imposing these so-called simplicity
constraints. They express that $B$ is the wedge product of a tetrad, restoring the degrees of
freedom of GR \cite{pleb}. The idea is thus to start with the standard spin foam model for BF
theory and to introduce the constraints at the quantum level. Their discretization involves
different subsimplices of the triangulated manifold, the most discussed being the so-called
diagonal and cross-simplicity constraints, respectively taking place at each triangle and each
tetrahedron. Efforts have thus concentrated on the quantization of the geometry of a tetrahedron
\cite{barbieri}, which is the natural arena for the Hamiltonian framework, using geometric
quantization \cite{baez barrett} to define a state space and quantum observables. According to the
symplectic structure, the variables discretizing the field $B$ are promoted to generators of
$\so(4)$ at the quantum level (see \cite{consistently solving, epr imm} typically). The constraints
can then be solved inducing restrictions on the spin data labelling the quantum states of the
tetrahedron. From this Hamiltonian perspective, the BC model strongly imposes the constraints. But
they form a second class system (see \cite{constraints classes} for a comparison of the different
models in this regard), and have thus to be weakly imposed to avoid the loss of degrees of freedom.
This is precisely the purpose of \cite{EPR}.

The issues are then to insert these results into the spin foam and to stick tetrahedra together.
This is usually done by identifying the spin data of the spin foam with the labels of the
tetrahedron state space, as in \cite{PRTVO} in the 3d case. Indeed, it enables to assign the same
spin to each copy of a triangle as a regluing process. Notice also that the amplitude of BC model
was built considering a single 4-simplex. From the point of view of \cite{FKLS}, the regluing of
tetrahedra is correctly performed in this model, but that of 4-simplices is failing, forgetting
that each tetrahedron is shared by two 4-simplices.

The standard way thus involves a recourse to both the Hamiltonian and the Lagrangian methods,
together with a gluing process which takes place at their meeting point. We here consider the
purely Lagrangian point of view, discretizing the functional integral. This approach has already
been explored to perturbatively define and compute generating functionals of theories written as a
BF term plus a (local) polynomial potential in $B$ \cite{BF action principle}. This work is in
particular based on the ability of performing the integration over the field $B$ in pure BF lattice
model. It remains unclear how such an integration can be exactly performed for a constrained BF
theory like \eqref{plebanski} (see however \cite{freidel conrady} for an interesting path integral
formulation of the new spin foam models, but not exactly based on the standard BF discretized
action). A second aspect in \cite{BF action principle} and \cite{freidel conrady}, which is
certainly related to the ability of performing the integration over $B$, is the breaking up of the
simplicial decomposition into disjoint 4-simplices. One thus consider an action for each 4-simplex
and compute its partition function in terms of boundary variables for the connection, without an
explicit regluing for $B$ (except the above-mentioned process based on the canonical analysis).

In this note, we provide an action for the BC model, which consists of a BF term supplemented with
the usual discretization of the simplicity constraints. This makes clear the physical setting of
the BC model. The BF part of the action is built on disjoint tetrahedra, and the constraints are
independently imposed on each of them. There is no explicit regluing of the variables $B$ assigned
to different copies of a triangle, but a gluing process with boundary variables for the connection
instead. Such a process is obviously sufficient for pure BF theory. However, we will show that the
Lagrange multipliers imposing the contraints behave as sources for the curvature, and make this
gluing fail to give the correct relations of parallel transport. The BC model has long been
suspected because of its lack of correlation between neighbouring simplices. It clearly appears
here that the key point in this regard is the gluing between tetrahedra once simplicity has been
taken into account for each of them. As a way-out in the spirit of Regge calculus, one may think of a  gluing involving the field $B$, before
its integration. We also provide a very simple parametrization of these correlations which could be
useful to get some flexibility to modify the BC model.

We clarify some issues concerning the BC model. Our parametrization of the constraints enables to
identify the physically relevant part of the partition function and
to use a freedom in the measure to show that the BC model weakly implements simplicity with specific fluctuations. The measure to be used is the most natural from the point of view of $\SU(2)$ lattice gauge theory. Another essential aspect is that the formulation \eqref{plebanski} contains gravity and also a
non-geometric sector. The two sectors are clearly identified and a spin foam model for the analog
of the BC model in the non-geometric sector is given.

The spin foam models of \cite{EPR, consistently solving, FKLS} intend to correct the BC model in
the view of different defects: it is supposed to strongly impose simplicity for \cite{EPR}, or to be built on
disjoint 4-simplices for \cite{FKLS}. The main feature of these models is the amplitude of a 4-simplex, based on
the 15j-symbol for Spin(4) with some fusion coefficients which supplant the BC intertwiner. The
amplitudes for triangles and tetrahedra however differ. The fusion coefficients are here shown to
appear in the framework of the BC model with a non-trivial measure. They are thus very natural
objects when trying to implement metricity and are not specific to the new models.

The organization of the paper is as follows. The first section reviews standard results of spin
foam quantization for Spin(4) BF theory. In the second section, we study the implementation of
diagonal simplicity in the functional integral. This is indeed usually considered as the starting
point to impose cross-simplicity, though it has never been precisely studied from this point of
view. This is the simplest modification of the BF spin foam model making it non-topological. We
tackle the issue of cross-simplicity in section \ref{sec BC}. The analysis is based on the
formulation of \cite{EPR}, and the BC model is derived from a discrete action principle. We use a
freedom in the measure to generalize the BC model in section \ref{sec BC measure}, and show how the
fusion coefficients of the new spin foam models naturally emerge in this setting. The measure is
also used to weaken the implementation of the constraints so as to recover the new spin foam
models. We close this note with a discussion concerning the physical setting of the BC model and
its improvement.


During the process of writing up the present paper, Conrady and Freidel \cite{freidel conrady,
freidel conrady 2} also provided a path integral formalism for the new Engle-Pereira-Rovelli and
Freidel-Krasnov spin foam models. Their approach is different from ours since the purpose of their
path integral is to exactly reproduce the amplitudes of these spin foam amplitudes, while our
formalism starts from the generic path integral for the discretized BF theory and explores how to
impose the simplicity constraints in a clear geometric way. At the end of their day, the present
work allows to derive a larger class of spinfoam models and provides them with a clear geometric
meaning while the Conrady-Freidel framework is more specific to the Engle-Pereira-Rovelli and
Freidel-Krasnov spin foam models and allows to extract their (semi-)classical behavior.

\section{Review of spin foams for BF theory and the discretized simplicity constraints} \label{review}

Let us first review some basics facts about the conventional spin foam model for topological
Spin(4) BF theory. We will all along use the splittings of the group, Spin(4)$=\SU(2)\times \SU(2)$
and of its algebra, $\spin(4) = \su(2)\oplus\su(2)$. Consider a triangulation of a 4-manifold. The
basis of Regge discretisation, in four dimensional spacetime, is to concentrate curvature on
2-simplices. Tetrahedra and 4-simplices are thus flat, and immmersed in the 4-manifold with
arbitrary, and so {\it a priori} different frames, that is local coordinates for the 'internal'
space, carrying the vectorial representation of Spin(4). The spin foams then lie on the 2-complex
dual to the triangulation. We will indifferently denote the triangles and their dual faces by $f$,
the tetrahedra and the edges of the dual faces by $\tau$, and the 4-simplices, dual to vertices, by
$v$.

\begin{figure} \begin{center}
\includegraphics[width=7.5cm]{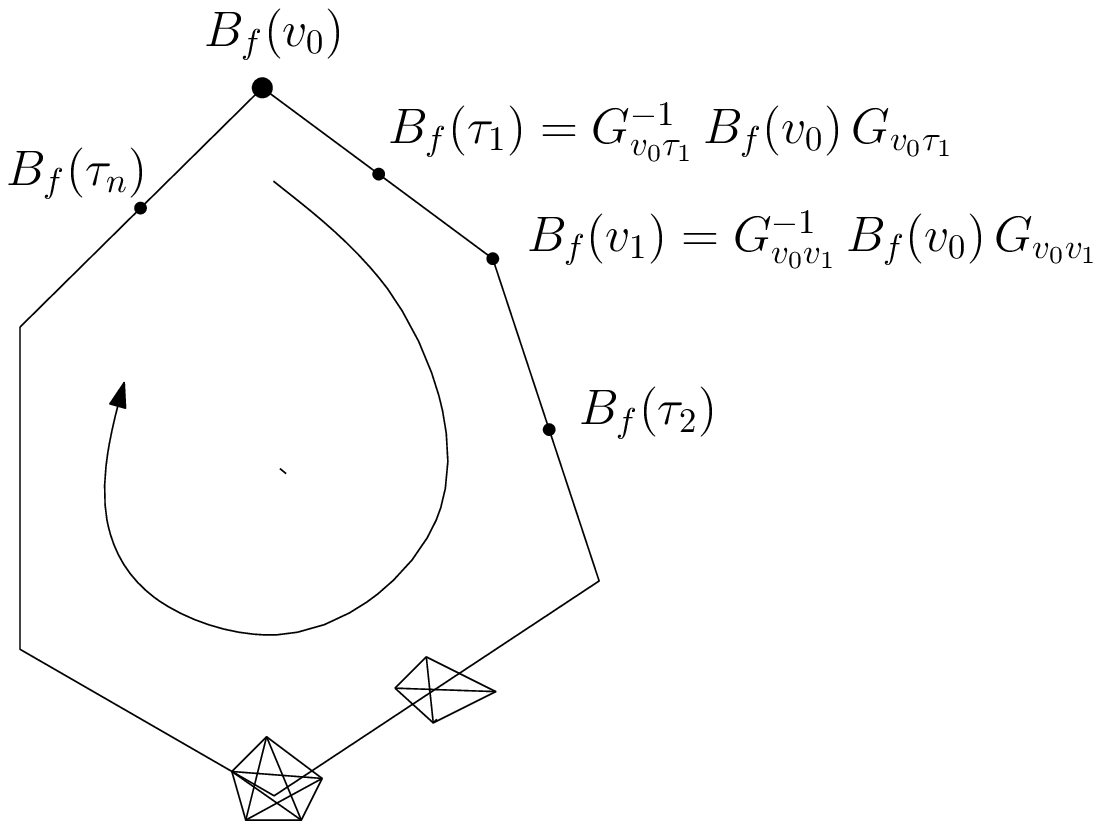}
\caption{ \label{bivectors fig} The dual face $f$ is equipped with an orientation and a base-point. The bivector $B_f(v_0)$ is defined in the frame of this base-point, and then parallelly transported along the boundary of $f$ until $B_f(\tau_n)$ is reached. This ensures that the action for BF theory does not depend on the choice of the base-points. }
\end{center}
\end{figure}

In the continuum, the theory is defined by the classical action $S_{BF} = \int \tr(B\wedge F)$,
where $B$ is a $\spin(4)$-valued 2-form, and $F$ the curvature of a Spin(4) connection. We define
the generators $\tau_i=-i\sigma_i/2$ of $\su(2)$, where the matrices $\sigma_i$ are the Pauli
matrices, which are anti-hermitian matrices satisfying $[\tau_i,\tau_j]=\eps_{ij}^{\phantom{ij}k}
\tau_k$. The trace '$\tr$' over $\su(2)$ is the conventional trace, with the normalisation:
$\tr(\tau_i\tau_j)=-\f{\delta_{ij}}{2}$. The discrete connection consists of Spin(4) group elements
$G_{v\tau}=(g_{+v\tau},g_{-v\tau})$ allowing parallel transport between frames along the boundary
edges of the dual faces. The orientation of each dual edge induces an orientation between each pair
of 4-simplices sharing a tetrahedron. The basic variables are, for each $\tau$, the holonomies
$G_{s(\tau)\tau}$ and $G_{\tau t(\tau)}$ representing the parallel transport between the 4-simplex
source $s(\tau)$ and $\tau$, and between $\tau$ and its 4-simplex target $t(\tau)$. Composing these
elements and given an orientation of each face, we define the holonomies around the
faces\footnotemark, $G_f(v_0)=G_{v_0\tau_1}G_{\tau_1 v_1} G_{v_1\tau_2}\cdots$, $v_0$ being a
base-point for $f$. Such a discrete holonomy can be transported in another frame of the boundary of
$f$, e.g. that of a tetrahedron, by the following obvious formula:
\footnotetext{
We suppose for the sake of notation that the orientation of each dual edge coincides with that of
the dual face $f$.}
\be \label{transport hol}
G_f(\tau) = G_{v_0\tau}^{-1}\, G_f(v_0)\, G_{v_0\tau}
\ee
where $G_{v_0\tau}$ turns around $f$ using the onwards path from $v_0$ to $\tau$. We discretize the
continuous 2-form $B$ on the triangles of the triangulation, with respect to the local frame of a
tetrahedron, $B_f(\tau)$, or that of a 4-simplex, $B_f(v)$. Each $B_f$ is a bivector, that is an
element of $\spin(4)$, $B_f = b_{+f}\oplus b_{-f}$, where each $b_{\pm f}$ can be seen both as a
$\su(2)$ matrix or a 3-vector in agreement with its definition: $b_{\pm} = b_{\pm}^i\tau_i$, with
$b_{\pm f}^i=\f{1}{2}(\f{1}{2}\eps^i_{\phantom{i}jk}B^{jk}\pm B^{0i})$. There is several ways to
consistently define the bivectors together with the rules for parallelly transporting them. A
natural way\footnotemark is to first define $B_f(v_0)$ for a base-point $v_0$, dual to a 4-simplex,
and then, using the orientation of $f$ and the elements $G_{\tau v}$, define the bivector in the
other frames:
\footnotetext{
Another method often used in spin foams is to divide dual faces into several pieces corresponding
to the choice of a boundary simplex. The so-called wedges correspond to pairs $(f,v)$.
Geometrically, it means that the initial triangulation has been broken up into disjoint simplices.
One can then define independent bivectors for each piece and reglue simplices in a specific way. It
will be detailed in the next sections, for we will argue that the natural setting of the BC model
corresponds to using pairs $(f,\tau)$.}
\be \label{def bivectors}
B_f(\tau_1) = G_{v_0\tau_1}^{-1}\, B_f(v_0)\, G_{v_0\tau_1},\qquad B_f(v_1) = G_{\tau_1v_1}^{-1}\, B_f(\tau_1)\, G_{\tau_1v_1}
\ee
and so on until $\tau_n$ is reached as in figure \ref{bivectors fig}. The relations
\eqref{transport hol} and \eqref{def bivectors} ensure that the discrete action for BF theory,
shown below, does not depend on the choice of a base-point for each dual face.

The Spin(4) gauge covariance reflects the arbitrariness in the choice of local orthonormal bases for the (flat) 4-simplices and tetrahedra. The gauge transformation $K$ thus acts through Spin(4) group elements $K_v$ and $K_\tau$ associated to 4-simplices and tetrahedra 
which rotate the local frames, as in the framework introduced in \cite{EPR}:
\begin{align}
K\ \triangleright\ &B_f(v) = K_v\ B_f(v)\ K_v^{-1} \label{gauge rule B} \\
K\ \triangleright\ &G_{v\tau} = K_{v}\ G_{v\tau}\ K_{\tau}^{-1}
\end{align}
The gauge invariant discretized BF action is then:
\be \label{BF action}
S_{BF} = \sum_f \Tr\Big(B_f(v_0)\, G_f(v_0) \Big) = \sum_f \tr\big(b_{+f}(v_0)\, g_{+f}(v_0)\big) + \tr\big(b_{-f}(v_0)\, g_{-f}(v_0)\big)
\ee
in which the traces are independent of the chosen frames thanks to \eqref{def bivectors}. It is also clear that \eqref{BF action} is invariant under the change $\tl{B}_f(v_0) = G_f(v_0)^nB_f(v_0)G_f(v_0)^{-n}$ for any $n$, since: $\tr(\tl{B}_f(v_0)G_f(v_0)) = \tr(B_f(v_0)G_f(v_0))$. This way of thinking avoids imposing that $B_f(v_0)$ is left invariant by $G_f(v_0)$. Such a relation, well-known in Regge calculus, comes in fact from the discretization of an equation of motion of Plebanski's action (we come back to this point in sections \ref{eom BC} and \ref{beyond BC}). Thus it may be sufficient to get it as e.o.m. for the discrete action principle as in \cite{freidel conrady 2} for the new spin foam models.

The spin foam model is then defined as the partition function for this action, and can be interpreted as a discrete way to compute the volume of the moduli space of flat connections. Indeed, the dynamical variables are the bivectors and the holonomies along the dual edges $G_\tau\equiv G_{s(\tau)\tau} G_{\tau t(\tau)}$, and the partition function reads:
\begin{align}
Z_{BF} &= \int \prod_\tau dG_\tau \int\prod_f dB_f\quad e^{i\sum_f \Tr\big(G_f(v) B_f(v)\big)} \\
 &= \int \prod_\tau dG_\tau\ \prod_f \delta\bigl( G_f\bigr) \label{flatness}
\end{align}
where $dB$ is the Lebesgue measure on $\spin(4) = \mathbb{R}^3\oplus\mathbb{R}^3$, and $dG=dg_+\,dg_-$ is the Haar measure on the group Spin(4)$=\SU(2)\times\SU(2)$. The action being linear in each $B_f$, their integrations lead to delta functions over the group for the holonomies around the dual faces\footnotemark. This is precisely the types of computations we want to perform for non-topological theories. In particular, the result of the integrations over the bivectors expressed as a function over the group gives a great part of the physical content of the theory. Here, it is a simple discretization of the condition asking for flatness of the connection \cite{baez} and, as in the continuum, bivectors are straightforwardly interpreted as Lagrange multipliers. Expanding the condition $\delta(G_f)$ according to the usual formula $\delta(g) = \sum_j d_j \chi_j(g)$ for $\SU(2)$ group elements $g$, where $d_j \equiv 2j+1$ is the dimension of the irreducible representation of spin $j$, the dual faces are labelled with representations $(j_{+f,},j_{-f})$:
\footnotetext{The integrals over $B_f$ is not altered when changing $g_{+f}$ or $g_{-f}$ into its opposite, so that the delta functions are in fact over $\SO(3)\times \SO(3)$ (see next section).}
\be \label{BF with rep}
Z_{BF} = \int \prod_\tau dG_\tau\ \sum_{\{(j_{+f,},j_{-f})\}} \prod_f d_{j_{+f}} d_{j_{-f}}\ \chi_{j_{+f}}\big(g_{+f}\big) \chi_{j_{-f}}\big(g_{-f}\big)
\ee

In the continuum, the action of general relativity can be written as that of BF theory supplemented with additional constraints which ensure that the field $B$ comes from the wedge product of a tetrad. They read:
\be \label{metricity}
\eps_{IJKL}\ B_{\mu\nu}^{IJ}\ B_{\lambda\sigma}^{KL} \propto \eps_{\mu\nu\lambda\sigma}
\ee
where $\mu,\nu,\lambda,\sigma$ are spacetime indices. In fact, the theory consists in two sectors, only one describing gravitation, which we will call the geometric sector, with $B^{IJ}=\eps^{IJ}_{\phantom{IJ}KL}\ e^K\wedge e^L$ for a tetrad 1-form $e$. The non-geometric sector, often given the misleading name of topological, is characterized by $B^{IJ} = e^I\wedge e^J$ (see \cite{pleb} for details). Its equations of motion only impose the vanishing of torsion and leave curvature free. The so-called simplicity constraints are conventionnally discretized on simplices of different dimensions:
\begin{align}
i)&\text{Diagonal simplicity constraints:}& &\eps_{IJKL}B_f^{IJ} B_f^{KL}=0,& &\text{for a single triangle} \label{diag simplicity}\\
ii)&\text{Cross-simplicity constraints:}& &\eps_{IJKL}B_f^{IJ} B_{f'}^{KL}=0,& &\text{for two triangles of a tetrahedron} \label{cross simplicity} \\
iii)&\text{Volume constraints:}& &\eps_{IJKL}B_f^{IJ} B_{f'}^{KL}\propto V_v,& &\text{for $f$ and $f'$ sharing a point in the 4-simplex $v$} \label{volume}
\end{align}
in which we have dropped the frame labels. The diagonal constraints express that the bivector of each face is simple, while the cross-simplicity constraints relate the bivectors of each tetrahedron. The third constraint \eqref{volume}, taking place at the level of each 4-simplex $v$, ask for the volume of $v$ to be independent of the bivectors used to compute it. It has been shown (see \cite{EPR} for instance) that \eqref{volume} can be replaced with the closure relation, $\sum_{f\subset\pp \tau} B_f(\tau) =0$ for each tetrahedron. The latter is usually taken for granted since it is an equation of motion of BF theory. It will also be the case here for the models under consideration. 

Thinking of adding the constraints to the BF action \eqref{BF action}, a natural idea is to use the fact that they are quadratic in the field $B$, so that Gaussian integrations can be performed. However, this does not turn out to be a great advantage in this context, even for the diagonal constraint \eqref{diag simplicity} which involves only one triangle. Indeed, expressions \eqref{diag simplicity}, \eqref{cross simplicity} and \eqref{volume} are not usual polynomial integrations in $B$, but constraints imposed with Lagrange multipliers instead. One has then to deal with determinants involving these Lagrange multipliers. This approach can be nevertheless fruitful in other contexts, for example in \cite{krasnov} which investigates the renormalization of Plebanski's action. A reformulation of the discretized constraints \eqref{diag simplicity} and \eqref{cross simplicity} has been introduced in \cite{EPR}, and used in \cite{consistently solving, FKLS, alexandrov}, which clarifies their geometric interpretation. It simply expresses that a tetrahedron whose triangles are labelled by bivectors satisfying the constraints spans a three dimensional subspace of Minkowski space (that is the 'internal space'), so that there exists for each tetrahedron a unit vector $N_\tau$ orthogonal to each of its faces in the following sense:
\be \label{new form simplicity}
\eps_{IJKL}\ N_\tau^{J} B_f^{KL} =0\qquad \text{or}\qquad N_{\tau J} B^{IJ}_f = 0,
\ee
respectively for the gravitational and non-geometric sectors. In the following sections, we will use another parametrisation, using $\SU(2)$ group elements, which can be straightforwardly interpreted in terms of \eqref{new form simplicity}.

\section{Simplicity of a single bivector} \label{sec simple rep}

In this section, we introduce basic ingredients from which non-topological spin foam models can be derived in a clean way, by looking in details at the diagonal constraints \eqref{diag simplicity}. First, it can be seen as an exercise to the challenge of imposing the cross-simplicity constraints. Our derivation of the BC model will be based on the same method. Second, the effects of the diagonal constraints have not been investigated so far from the point of view of the lattice path integral. The physical content of \eqref{diag simplicity} is that each $B_f$, or equivalently their Hodge dual, $\star B_f^{IJ}=\f{1}{2}\eps^{IJ}_{\phantom{IJ}KL} B_f^{KL}$, is simple, that is to say that it is an anti-symmetrized product of vectors, $u^{[I} v^{J]}$. It thus corresponds to a genuine non-topological model.

The diagonal constraints \eqref{diag simplicity} are usually imposed thanks to a geometric interpretation which identifies the Spin(4) representations $(j_{+f},j_{-f})$ labelling faces in the state-sum model \eqref{BF with rep}  for BF theory, with a quantization of the triangle areas. Such an identification is motivated by the canonical analysis of the discrete action. It then leads to implementing the diagonal constraints by imposing the equality of the self-dual and anti-self-dual representations $j_{+f}=j_{-f}$, which corresponds to the simple representations of Spin(4). Here, we show in a rigourous way to what extent this implementation is correct.

The usual imposition of the constraints \eqref{diag simplicity}, which borrows tools from geometric quantization, leads to the following partition function:
\begin{align}
Z_{\mathrm{simple}} &= \int \prod_\tau dG_\tau\ \prod_f Z_f(G_f), \\
\text{where}\quad
Z_f(G_f) &= \sum_{j_f} d_{j_f}^k\ \chi_{j_f}\bigl(g_{+f}\bigr)\ \chi_{j_f}\bigl(g_{-f}\bigr) \label{Z_f simple rep ansatz}
\end{align}
with $k=2$ defining the usual face amplitude. We will show how the discrete path integral for the BF action with the diagonal constraints naturally gives \eqref{Z_f simple rep ansatz} with $k=0$, that is a trivial face amplitude, and how this result generalizes to:
\be \label{ansatz simple rep measure}
Z_f(G_f) = \chi\big(g_{+f}\big)\ \sum_{j_f} \chi_{j_f}\bigl(g_{+f}\bigr)\ \chi_{j_f}\bigl(g_{-f}\bigr)
\ee
where $\chi$ is a $\SU(2)$ class function, typically a character, to ensure the gauge invariance of $Z_f$. $\chi$ is a measure ambiguity of the path integral, and as we will see, the models defined by \eqref{ansatz simple rep measure} share the same basic physical features due to \eqref{diag simplicity}. When $\chi$ is a character, it induces simple shifts of the spin of the self-dual character. The reason why \eqref{ansatz simple rep measure} does not break the symmetry between the self-dual and anti-self-dual sectors will be clear below.

Since our aim is to explicitly perform the integrals over the bivectors in the partition function, the key point is to find a convenient parametrization of the solutions to \eqref{diag simplicity}. Decompose the bivectors $B_f$ into self-dual and anti-self-dual parts: $B_f=\vec{b}_{+f}\oplus\vec{b}_{-f}$, with $b_{\pm f}^i$ given above. The diagonal simplicity constraint is then very easy to deal with. It reads: $\lvert \vec{b}_{+f}\rvert^2=\lvert \vec{b}_{-f}\rvert^2$, that is the self-dual and anti-self-dual parts of $B_f$ have equal norm. Thus, there exists a $\SU(2)$ rotation $h_f$ which maps one into the other:
\be \label{diag sol}
b_{-f}=-h_f^{-1}\ b_{+f}\ h_f
\ee
written in the adjoint representation. Since the bivectors are defined in specific frames, so are the rotations, $h_f(v)$ or $h_f(\tau)$. As explained in \cite{consistently solving}, the rotation $h_f$ has the following interpretation. The Spin(4) group element $H_f=(h_f,\mathrm{id})$ maps the vector $N^{(0)}=(1,0,0,0)$ into a unit vector $N_f$ orthogonal to the Hodge dual of $B_f$, $\eps_{IJKL}\ N_f^J B_f^{KL}=0$. Note that there is a sign ambiguity: we could have simply chosen $b_{-f}=h_f^{-1}b_{+f}h_f$. Then, the vector $N_f = H_f N^{(0)}$ is orthogonal to $B_f$. This ambiguity is related to the fact that the simplicity constraints are solved in the continuum by two different sectors. The geometric one corresponds to the solution $B=\star(e\wedge e)$ for a tetrad 1-form $e$, while the non-geometric one corresponds to $B=e\wedge e$. As shown in \cite{consistently solving}, the sign ambiguity distinguishes between these two sectors, although it is irrelevant as far as one only deals with diagonal simplicity. In the following sections, we will be able to identify the two sectors at the quantum level, since we will only have to identify them at the classical level in the path integral. Our analysis will be done for the geometric sector all along the paper, except when it is interesting to compare expressions between the two sectors, in which case it will be explicitly stated.

Instead of imposing \eqref{diag sol} with delta functions, we add them to the action with $\SU(2)$ Lagrange multpliers $q_f$:
\be \label{simple rep action}
S_{\mathrm{diag}} = \sum_f \Tr\Bigl(B_f(v) G_f(v)\Bigr) + \tr\Bigl[q_f(v) \bigl(b_{-f}(v) + h_fb_{+f}h_f^{-1}(v)\bigr)\Bigr]
\ee
The choice of $q_f\in\SU(2)$ is the natural choice since the equations of motion obtained by
varying \eqref{simple rep action} with respect to $b_{-f}$ and $b_{+f}$ clearly relate $g_{-f}$ and
$g_{+f}$ to $q_f$ and $h_f q_f h_f^{-1}$. Since the action is linear in the bivectors, integrating
them projects the fields to these equations of motion, so they will be detailed just below. We
allow for a general measure $\mu$ for the Lagrange multipliers, thus taking into account different
possible weights for the quantity $(b_{-f}+h_f^{-1}b_{+f}h_f)$:
\begin{align} \label{simple rep1}
Z_{\mathrm{diag},f,\mu} &= \int dB_f\ e^{i\mathrm{tr}(B_f G_f)}\ \int dh_f\ \tl{\delta}_\mu\bigl(b_{-f}+h_f^{-1}b_{+f}h_f\bigr) \\
\text{with}\quad \tl{\delta}_\mu(x_f) &= \int_{\SU(2)} dq_f\ \mu(q_f)\ \exp\big\{i\mathrm{tr}[q_fx_f]\big\}
\end{align}
Because of the gauge transformation properties of $(b_{-f}+h_f^{-1}b_{+f}h_f)$, $h_f$ transforms
with the self-dual $\SU(2)$ on the left and the anti-self-dual $\SU(2)$ on the right. $q_f$
transforms under the adjoint representation of the anti-self-dual sector, and $\mu$ is then taken
to be a class function. It is easy to see\footnotemark\ that the strong imposition of the
constraints, that is $\tl{\delta}_\mu=\delta$, is realized for $\mu(q) = \lvert \tr\ q\rvert$. We
will find a precise relation between $\mu$ and the function $\chi$ of \eqref{ansatz simple rep
measure}.

\footnotetext{
The Haar measure over $\SU(2)$ can be related to the Lebesgue measure over $\mathbbm{R}^3$ with the
help of a factor $\lvert \tr q\rvert$ taking into account the structure of the multiplication law
of $\SU(2)$ and its compactness (see appendix \ref{formula}). Strictly speaking, it is not true
that we have in this case $\tl{\delta}_\mu=\delta$ since the domain of integration remains compact.
One can see however that it has the same effect, by first integrating the bivectors. In particular,
the trivial weight here gives: $\int dg\ e^{i\tr(bg)} = \f{J_1(\lv b\rv)}{\lv b\rv}$ whose maximum
value is reached for $\lv b\rv =0$, $J_1$ being the Bessel function of the first kind of order 1,
up to irrelevant coefficients.}

The integral over the bivector $B_f$ in \eqref{simple rep1} can easily be performed with the help of \eqref{int b}:
\be \label{simple rep after b int}
Z_{\mathrm{diag},f,\mu} = \f{1}{\lvert \tr(g_{+f})\ \tr(g_{-f})\rvert}\ \int dh_f dq_f\ \mu(q_f)\
\delta\big(g_{-f}\, q_f\big)\ \delta\big(g_{+f}\, h_f\, q_f\, h_f^{-1}\big)
\ee
The delta functions in this equation are in fact over $\SO(3)$ and not $\SU(2)$\ \footnotemark.
Thus, the Lagrange multipliers $q_f$ stand for the holonomies and relate the anti-self-dual part
$g_{-f}$ to the self-dual part $g_{+f}$ conjugated by $h_f$:
\footnotetext{
Indeed, the expression $\int d^3b\ \exp(i\tr(bg))$ does not distinguishes between $g$ and $-g$ (see
appendix \ref{formula}). Thus, the expansion into characters to be used in the following is:
$\delta_{\SO(3)}(g) = \sum_{j\in\mathbbm{N}}d_j \chi_j(g)$, and all irreducible representations are
of integral spin.}
\be
Z_{\mathrm{diag},f,\mu} = \f{\mu(g_{-f})}{\lvert \tr(g_{+f})\ \tr(g_{-f})\rvert}\ \int dh_f\
\delta\big(g_{+f}\, h_f\, g_{-f}^{-1}\, h_f^{-1}\big) \label{physical step simple rep}
\ee
Since \eqref{physical step simple rep} imposes $g_{+f}$ and $g_{-f}$ to be related by group
conjugation, it is clear that they share the same class angle. Since $\mu$ is a class function, we
have $\mu(g_{-f})=\mu(g_{+f})$. The key quantity in \eqref{physical step simple rep}, responsible
for the non-topological character of the model, is obviously $\delta(g_{+f} h_f g_{-f}^{-1}
h_f^{-1})$, which peaks the amplitude on classical configurations. Notice indeed that the group
element $g_{+f} h_f g_f^{-1}$ is just the result of parallelly transporting the rotation $h_f$
around the dual face $f$. The amplitude thus selects holonomies which preserve the normal vector
$N_f$ defined by $h_f$ for each face. This restriction on the holonomy degrees of freedom can be
seen as a characterization or a definition of the model, exactly as the BF action serves as an action principle for
the model defined by the flatness condition, $\delta(G_f)$ \eqref{flatness}.

Using the usual expansion $\delta(g) = \sum_j d_j \chi_j(g)$ and the Peter-Weyl theorem, the
integrations over the rotations simply lead to:
\be \label{final simple rep}
Z_{\mathrm{diag},f,\mu} = \f{\mu(g_{+f})}{\left(\tr \,g_{+f}\right)^2}\ \sum_{j_f}
\chi_{j_f}\bigl(g_{+f}\bigr)\
\chi_{j_f}\bigl(g_{-f}\bigr)
\ee
which is precisely \eqref{ansatz simple rep measure} with measure factors related by: $\mu(g) =
(\tr\,g)^2\chi(g)$. Thus, to get rid of them in the final expression, we have to choose $\mu(g) =
(\tr\,g)^2$, which corresponds to imposing the diagonal constraints, not with a delta, but with:
\be \label{bessel measure}
\chi=1\ \longleftarrow\ \tl{\delta}_\mu(\vec{x}) = \f{J_1(\lvert \vec{x}\rvert)+J_3(\lvert \vec{x}\rvert)}{\lvert \vec{x}\rvert}
\ee
instead, up to irrelevant factors. Here, $J_1$ and $J_3$ are Bessel functions of the first kind
defined by: $J_n(x) = \f{1}{\pi i^n}\int_0^\pi e^{ix\cos \theta}\ \cos(n\theta) d\theta$. The
function $(J_1(x)+J_3(x))/x$ is peaked around 0, but with a finite width. This means that some
fluctuations around the solutions of the diagonal constraints are allowed. These fluctuations are
such that that integrating over the bivectors $B_f$ yields true delta functions over the group in
\eqref{physical step simple rep}. Since we build spin foam models from group integrals, this measure is a natural choice.

More generally, inserting a character $\mu(g) = \chi_l(g)$ in the representation of spin $l$ can be
translated into the choice:
\be
\tl{\delta}_{\chi_l}(\vec{x}) = \f{J_{2l+1}(\lvert \vec{x}\rvert)}{\lvert \vec{x}\rvert}
\ee
Such a function is not peaked around 0 anymore (it even vanishes at that point for all $l>1$).
Instead, it has a maximum which is reached for a value of $x$ depending on $l$. The fact that we
nevertheless obtain the BF theory restricted to simple representations of Spin(4) can be traced
back to the special form of $\tl{\delta}_\mu$, given as a Fourier transform evaluated at
$b_{-f}+h_f^{-1}b_{+f}h_f$. Then, upon integrating the bivector, the field configurations are
always restricted to those satisfying $g_{+f} = h_f g_{-f} h_f^{-1}$.

Let us introduce a new way to insert the constraints into the discrete path integral. This way of
proceeding will be argued in this work to be very natural, and will give a very clean action
principle for the BC spin foam model. Since equation \eqref{simple rep after b int} enforces the
$\SU(2)$ variables $q_f$  to be the holonomies, the constraints should be better introduced into
the action in a way which respects the group structure and avoids the measure factors $(\tr\,
g_{-f})^2$. Indeed, the constraints have been naively added to the action in \eqref{simple rep action}, so that the latter involves sums of group elements like: $\tr[b_{-f}(g_{-f}+q_f)]$. The measure factors $\lv \tr g\rv$ in \eqref{simple rep after b int} are directly due to summing group elements instead of using the natural $\SU(2)$ multiplication law (see \eqref{int b}), and do not have any clear meaning. The adapted tool is thus a non-commutative addition $\oplus$ which multiply group elements instead of summing them:
\be \label{nc law}
e^{\tr(bg)\oplus \tr(bh)}\equiv e^{\tr(bgh)} \ne e^{\tr(b(g+h))}
\ee
It deforms the composition of plane waves when the momentum space is given by the group $\SU(2)$. This new operation is
naturally compatible with the $\SU(2)$ group multiplication and removes in a natural way all weird measure factors. More precisely, we will show that it implicitly takes into account the measure \eqref{bessel measure}. It is quite convenient when working
with lattice gauge theories involving elements from both a Lie group and it Lie algebra. In particular, it was already shown to arise when
looking at observables in the Ponzano-Regge spinfoam model for 3d quantum gravity \cite{EQFT}.

The diagonal simplicity constraints can be implemented at the level of the action with the
following form:
\begin{align}
S_{\star,\mathrm{diag}} &= \sum_f \Tr\Bigl(B_f(v)\, G_f(v)\Bigr) \oplus \tr\Bigl[q_f(v) \bigl(b_{-f}(v) + h_f^{-1}\,b_{+f}\,h_f(v)\bigr)\Bigr] \\
 &= \sum_f \tr\bigl(b_{-f}(v)\, g_{-f}(v)\, q_f(v)\bigr) + \tr\bigl(b_{+f}(v)\, g_{+f}(v)\, h_f\,q_f\,h_f^{-1}(v)\bigr) \label{diag action}
\end{align}
Written in this form, the action can be interpreted as a BF theory with 'bivectors'\footnotemark\ $\tl{B}_f$ of a special form:
\footnotetext{
The term 'bivectors' is here a misuse of language. Indeed, the following $\tl{b}_{\pm f}$ do not
live in $\su(2)$ anymore. We here simply want to stress that the action gets a simple form, as a BF
theory with bivectors modified, or deformed, by the constraints.}
\be
\tl{B}_f(v) = Q_f(v)\,B_f(v),\qquad \qquad \text{with}\qquad Q_f(v) = \bigl(h_f\,q_f\,h_f^{-1}(v),q_f(v)\bigr)
\ee

This action can be easily seen to reproduce the previous computations, but avoiding the measure
factors $(\tr\, g_{-f})^2$. Indeed, the integrations over $b_{-f}$ and $b_{+f}$ only need the
formula
\eqref{int b} with one of the two group elements being the identity. One has instead of
\eqref{simple rep after b int}:
\be \label{content simple rep}
Z_{\star,\mathrm{diag},\mu} = \int \prod_\tau dG_\tau \prod_f Z_{\star,\mathrm{diag},f,\mu}\quad\text{with} \left\{
 \begin{split}
 Z_{\star,\mathrm{diag},f,\mu} &= \int dh_f\ dq_f\ \mu(q_f)\ \delta\big(g_{-f}\, q_f\big)\ \delta\big(g_{+f}\, h_f\, q_f\, h_f^{-1}\big) \\
 &= \mu(g_{-f})\ \int dh_f\ \delta\big(g_{+f}\, h_f\, g_{-f}^{-1}\, h_f^{-1}\big).
\end{split} \right. \ee
so that \eqref{Z_f simple rep ansatz} is recovered with the face weight $k=0$ and the choice
$\mu=1$. The sum $\oplus$ can thus be said to naturally impose the constraints in a weak manner
which is adapted to spin foams, the measure \eqref{bessel measure} being taken into account through
the $\SU(2)$ group structure.

Let us come back to the sign ambiguity of \eqref{diag sol}. Having chosen $b_{-f}+h_f^{-1} b_{+f}
h_f =0$, the field restrictions obtained in \eqref{content simple rep} through the BF action are:
$g_{+f}h_fg_{-f}^{-1}h_f^{-1}=\mathrm{id}$. Thus, the sign $+$ between $b_{+f}$ and $b_{-f}$ has
been translated into taking the inverse of $g_{-f}$. This duality gives for the other sector:
\be
b_{-f} - h_f^{-1}\ b_{+f}\ h_f = 0 \qquad \stackrel{\int dB_f}{\longrightarrow}\qquad
\delta\Big(g_{+f}\ h_f\ g_{-f}\ h_f^{-1}\Big) \label{topo sector simple rep}
\ee
In this case, the orthogonal vectors $N_f$ are left invariant by the group elements $\star G_f =
(g_{+f},g_{-f}^{-1})$. The duality between the constraints on the bivectors and the restrictions on
the holonomies can be enlightened by using the normals $N_f$ instead of the rotations $h_f$.
Writing the BF action, for a face $f$, $\Tr(B_f G_f)= B_f^{IJ} P_{f,IJ}$, we introduce a Lagrange
multiplier $\lambda_f$ such that:
\be
Z_{\mathrm{diag},f} = \int dN_f \int dB_f \int d\lambda_f\ e^{i B_f^{IJ} P_{f,IJ} + i\lambda_f\ (\star B_f)^{IJ} N_{f,J} (\star B_f)_{IK} N_f^K}
\ee
Thanks to the Spin(4) gauge covariance, it is sufficient to analyse this integral for
$N_f=N^{(0)}$. In this case, we have $B_f^{ij}=0$ for $i,j=1,2,3$, so that only the components
$B_f^{0i}$ are free. Integrating the BF action then implies that $P_f^{0i}=0$, or in covariant
form: $N_{fJ} P_f^{IJ}=0$. Thus:
\be
Z_{\mathrm{diag},f} = \int dN_f\ \delta\Big(N_{f J}\, P_f^{IJ}\Big)
\ee
which is equivalent to \eqref{content simple rep}. In contrast, imposing instead $N_{f J}
B_f^{IJ}=0$ leads to the restriction: $\eps_{IJKL}\, N^J_f P_f^{KL} = 0$. However, when only
dealing with the diagonal constraints, the sign ambiguity is irrelevant, as one can check by
integrating over $h_f$ the r.h.s. of \eqref{topo sector simple rep}.

\section{An action principle for the Barrett-Crane model} \label{sec BC}

\subsection{Imposing cross-simplicity}

Inserting the solutions $B_f=(b_{+f},-h_f^{-1}b_{+f}h_f)$ of the diagonal constraints derived in the previous section into the cross-simplicity constraints \eqref{cross simplicity} gives, in vectorial notations:
\be \label{simplicity}
\vec{b}_{+f}\cdot \vec{b}_{+f'} = \vec{b}_{+f}\cdot R(h_f\,h_{f'}^{-1})\,\vec{b}_{+f'}
\ee
for pairs of triangles $(f,f')$ in the boundary of a tetrahedron. Note that the sign ambiguity is here irrelevant, as expected since the sector of the theory is selected by the choice of a sign when imposing (diagonal) simplicity for all bivectors. The constraints \eqref{simplicity}  relate the rotations $h_f$ for each pair of adjacent triangles, so that a clear geometric interpretation is expected to emerge at the tetrahedron level. Indeed, in \cite{EPR, FKLS, consistently solving}, the constraints are imposed into the spin foam model of BF theory by considering that there exists a vector $N_\tau$ orthogonal to each tetrahedron $\tau$ (precisely, to the Hodge dual of all bivectors of $\tau$). Here, in terms of the rotation variables, the cross-simplicity constraints are easily solved by : $h_f\ h_{f'}^{-1} = \phi_f\ \phi_{f'}^{-1}$, where $\phi_f$ and $\phi_{f'}$ are elements of the $\U(1)$ subgroups respectively preserving $\vec{b}_{+f}$ and $\vec{b}_{+f'}$. However, since we are solving \eqref{simplicity} in a given tetrahedron $\tau$, the $\U(1)$ elements depend {\it a priori} on $\tau$. The one-parameter family which solves the constraints for a given face can be thus be parametrized by:
\be \label{solv simplicity}
b_{-f} = - h_f^{-1}\ b_{+f}\ h_f,\qquad \text{with}\quad h_f = \phi_{f,\tau}\,h_\tau\quad\text{and}\quad \phi_{f,\tau}\ b_{+f}\ \phi_{f,\tau}^{-1} = b_{+f}
\ee
where $h_\tau$ is taken to be the same for all triangles of $\tau$, and is most naturally defined in the frame of $\tau$. As the rotations $h_f$ do not depend on $\tau$, there exists a relation between $h_{\tau_1}$ and $h_{\tau_2}$ when $\tau_1$ and $\tau_2$ share a face $f$:
\be \label{coupling tetra}
h_{\tau_1}(v)\ h_{\tau_2}^{-1}(v) = \phi_{f,\tau_1}^{-1}(v)\ \phi_{f,\tau_2}(v)
\ee
This relation holds when $h_{\tau_1}$ and $h_{\tau_2}$ are expressed in the same frame. The geometric meaning is the following. A triangle in 4d admits an orthogonal plane, defined by unit vectors $N$ satisfying $\eps_{IJKL}\,N^J B_f^{KL} = 0$. On this plane, $h_f$ stands for the choice of a direction $N_f = (h_f, \mathrm{id}) N^{(0)}$. One can check that in turn $h_\tau$, defined by \eqref{solv simplicity}, $h_\tau = \phi_{f\tau}^{-1}\,h_f$, defines a unit vector $N_\tau = (h_\tau, \mathrm{id}) N^{(0)}$ orthogonal to each face of $\tau$, that is $\eps_{IJKL}\ N_\tau^J B_f^{KL}(\tau) = 0$. Notice that these (classical) relations correspond to the geometric sector. The fact that $h_f$ and $h_\tau$ are related by a $\U(1)$ element simply expresses that $N_f$ and $N_\tau$ lie in the same plane orthogonal to $f$. Similarly, when $\tau_1$ and $\tau_2$ share the face $f$, the rotations $h_{\tau_1}$ and $h_{\tau_2}$ are not independent because the normal vectors $N_{\tau_1}$ and $N_{\tau_2}$ are both constrained to lie in a well-defined plane, that orthogonal to $\star B_f$.

A gauge transformation acts on $h_\tau$ as:
\be \label{gauge rule h}
K\ \triangleright\ h_{\tau} = k_{+\tau}\ h_{\tau}\ k_{-\tau}^{-1} \\
\ee
This specific rule implies that we can always choose a gauge in which $h_\tau = \mathrm{id}$, and thus $N_\tau=N^{(0)}$ being orthogonal to all triangles of $\tau$. Notice the difference with the rotations $h_f$ which only impose simplicity for each single bivector. Such a rotation can be gauged fixed to the identity in a given frame, say that of a tetrahedron which has $f$ in its boundary. However, it is then impossible to fix the rotations $h_f$ of the other faces of $\tau$ to the identity. From this point of view, the $\U(1)$ elements $\phi_{f\tau}$ enable geometric consistency between gauges of frames and normals to triangles.

Notice that the relation $b_{-f}(\tau)=-h_\tau^{-1} b_{+f}(\tau) h_\tau$ implicitly contains the elements $\phi_{f,\tau}$, as far as a single bivector is associated to each face, so that they can be trivially integrated out. This parametrization of the constraints is introduced in the path integral measure:
\be \left. \begin{split}
\int &\prod_f d^3b_{+f} d^3b_{-f}\ \delta\bigl(\lvert b_{+f}\rvert^2 - \lvert b_{-f}\rvert^2\bigr) \\
&\times\prod_\tau \prod_{(f,f')\subset\pp\tau} \delta\bigl( \vec{b}_{+f}\cdot\vec{b}_{+f'} - \vec{b}_{-f}\cdot\vec{b}_{-f'}\bigr) \end{split}\right\}
\quad\longrightarrow\quad
\int \prod_f d^3b_{+f} d^3b_{-f}\ \int \prod_\tau dh_\tau \prod_{(f,\tau)} \delta\big(b_{-f}(\tau)+h_\tau^{-1}\, b_{+f}(\tau)\, h_\tau\big)
\ee
This is in contrast with what can be read in the literature. Indeed, the spin foam models trying to implement the cross-simplicity constraints are usually built up only imposing them tetrahedron by tetrahedron. This means that the triangulation has been initially broken up into a disjoint union of tetrahedra, so that there are as many copies of a given triangle as there are tetrahedra sharing it. A bivector $B_{f\tau}$ is associated to each copy of a face $f$, one per tetrahedron having $f$ in its boundary, and they are all taken to be independent of each other. This corresponds to the following measure:
\be \label{broken up measure}
\int \prod_\tau \prod_{f\subset\pp\tau} d^3b_{+f\tau} d^3b_{-f\tau} \int \prod_\tau dh_\tau \prod_{\substack{\tau \\ f\subset\pp\tau}} \delta\bigl(b_{-f\tau}+h_\tau^{-1}\,b_{+f\tau}\,h_\tau\bigr)
\ee
We will focus on this specific situation for the rest of the section. However, an essential drawback of this approach already appears. The measure \eqref{broken up measure} does not encode the relations \eqref{coupling tetra} between adjacent tetrahedra. Indeed, the vector $N_\tau$ defined by $h_\tau$ is only orthogonal to the Hodge dual of the bivectors $B_{f\tau}$, and not to the bivectors labelling the other copies of the triangle $f$. This measure will be supplemented in the following with an action naturally built on disjoint tetrahedra and a regluing process using boundary variables for the connection, as in \cite{BF action principle}, leading then to the BC model. Since the relations \eqref{coupling tetra} are not imposed at the quantum level, it will be interesting to know whether or not the regluing is sufficiently efficient to make them hold classically. This issue is not clear since we will use boundary variables for the connection while the curvature degrees of freedom will not be frozen (unlike in BF theory).

It is not surprising that the BC model implements the constraints in this simplified way since an underlying idea is the geometric quantization of a single tetrahedron \cite{barbieri, baez barrett}. As explained above, the same can be said for the new vertices. We can also think that restricting attention to the space of intertwiners associated to a single tetrahedron in order to solve the constraints, the correlations between tetrahedra may be reduced with respect to \eqref{coupling tetra}. The point is naturally that of the gluing process consisting in assigning the same $\SU(2)$ representation $j_f$ to every copy of each face.

\subsection{A derivation of the BC model: Imposing the constraints on disjoint tetrahedra}

We now propose to study the simplified situation most often considered in the literature, corresponding to \eqref{broken up measure}. The aim is obviously to provide a clean link between an action principle and the derived spin foam model, by explicitly performing the integrals over the bivectors. It has been argued that strongly imposing the constraints leads to the BC model, and that the latter has no equivalent in the other sector of the theory, while the new vertices weakly implement the constraints. The main result of this section is that the most naturally obtained spin foam model in this framework is the BC model. It in fact corresponds to a weak imposition of the constraints, more precisely to the choice of the measure \eqref{bessel measure} allowing specific fluctuations around the solutions of the constraints. Moreover, this clean treatment shows that the BC model is indeed a quantization of the geometric sector, and that its equivalent in the non-geometric sector is a spin foam model for BF theory restricted to the simple representations of Spin(4), with specific face and tetrahedron amplitudes.

\begin{figure} \begin{center}
\includegraphics[width=5cm]{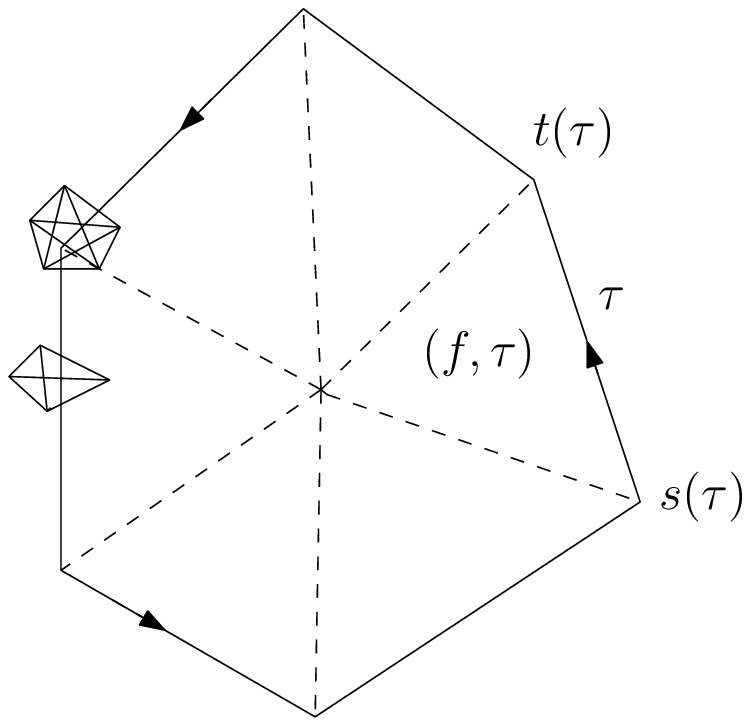} \qquad\qquad\qquad \includegraphics[width=4.5cm]{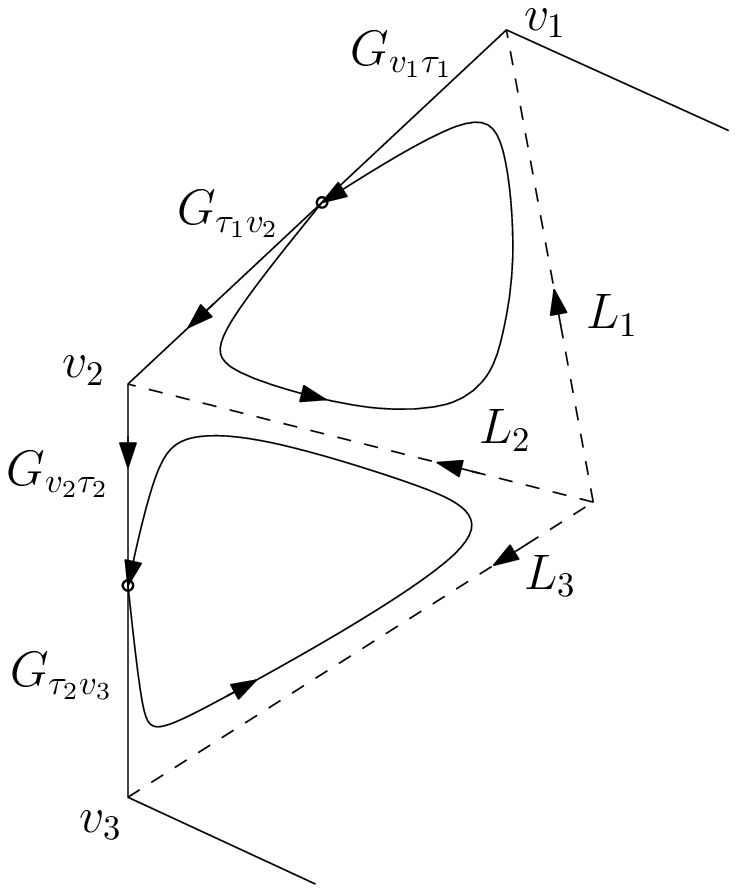}
\caption{ \label{variables Bftau} The left picture represents a dual face, whose edges, denoted $\tau$, are dual to tetrahedra and vertices dual to 4-simplices. The source and target vertices of $\tau$ are respectively denoted $s(\tau)$ and $t(\tau)$. The dashed lines delimit the pairs $(f,\tau)$ on which we define the bivectors $B_{f\tau}$ and which correspond to the triangles of a triangulation initially broken up into tetrahedra. The 'holonomies' around the pairs $(f,\tau)$ are defined on the right picture, the base-points being the tetrahedra. The internal links carry group elements $L_{fv}$, each shared by exactly two tetrahedra. They can be seen as boundary connection variables for each $\tau$, and integrating them is thought of as a regluing of tetrahedra all together. We have: $G_{f\tau} = G_{\tau\, t(\tau)}\, L^{-1}_{ft(\tau)}\, L_{fs(\tau)}\, G_{s(\tau)\, \tau}$.}
\end{center}
\end{figure}

Since we want to impose the constraints independently for each tetrahedron, we assign independent bivectors for each triangle of each tetrahedron. This can be achieved by dividing the dual faces into pieces such that each piece, denoted $(f,\tau)$, is completely specified by a dual edge of the face. This means that the triangulation has been initially broken up into a disjoint union of tetrahedra. A copy of a triangle is thus specified by a tetrahedron having it in its boundary. These variables are very similar to the wedge variables which instead consist of pairs $(f,v)$. Like with the wedges, one can build the action for BF theory by coupling the bivectors $B_{f\tau}$ to group elements $G_{f\tau}$ (see figure \ref{variables Bftau} and appendix \ref{naive BC} for the details of the construction). The bivectors $B_{f\tau}$ are defined in the tetrahedron frames. Important ingredients are Spin(4) elements $L_{fv}$ assigned to the internal edges dividing $f$. They can be seen as boundary variables for the connection, which glue tetrahedra in a certain way discussed in the next section.

Our starting point is an action for pure BF theory:
\be \label{broken up action}
\sum_{(f,\tau)} \Tr \big( B_{f\tau}\, G_{f\tau}\big)
\ee
in which the boundary variables $L_{fv}$ really reglue the pieces $(f,\tau)$ of each dual face, expressing the vanishing of the curvature. In this broken up triangulation picture, we want to impose the constraints encoded in \eqref{broken up measure}:
\be \label{cross-simplicity BC}
b_{-f\tau} = -h_\tau^{-1}\ b_{+f\tau}\ h_\tau
\ee
which state the existence of independent unit vectors $N_\tau$ orthogonal to tetrahedra, or more precisely, $\eps_{IJKL}B_{f\tau}^{JK} N_\tau^L=0$. Notice that equation \eqref{cross-simplicity BC} is completely classical and corresponds to the geometric sector. As far as integration over the bivectors is concerned, we can use the results of section \ref{sec simple rep} simply replacing the labels $f$ with $f,\tau$ except for $h_f$ which becomes $h_\tau$. Thus, gathering \eqref{broken up action} and \eqref{cross-simplicity BC} with Lagrange multipliers $q_{f\tau}\in\SU(2)$, we form the action:
\be \label{BC action}
S^{(0)}_{BC} = \sum_{(f,\tau)} \Tr \bigl( B_{f\tau}\, G_{f\tau}\bigr) + \Tr \bigl[q_{f\tau}\, (b_{-f\tau}+ h_\tau^{-1}\, b_{+f\tau}\, h_\tau)\bigr]
\ee
It is shown in appendix \ref{naive BC} that this action leads to the BC model, using the measure of finite width \eqref{bessel measure} for the bivectors $B_{f\tau}$. As explained in section \ref{sec simple rep}, this measure is related to the fact that group elements are summed in the action \eqref{BC action}, like: $\tr[b_{-f\tau}(g_{-f\tau}+q_{f\tau})]$. The specific measure \eqref{bessel measure} can be dropped if we use the natural group multiplication instead, i.e. the composition $\oplus$ \eqref{nc law} for each piece $(f,\tau)$, which implicitly takes it into account. This corresponds to the following action:
\be \label{BC action 1}
S^{(1)}_{BC} = \sum_{(f,\tau)} \tr\bigl(b_{-f\tau}\,g_{-f\tau}\,q_{f\tau}\bigr) + \tr\bigl(b_{+f\tau}\,g_{+f\tau}\,h_\tau q_{f\tau}h_\tau^{-1}\bigr)
\ee

We have nevertheless noticed that a simpler action leads to the BC model in a very natural way. The underlying idea, to say it roughly, is to stick the pieces $(f,\tau)$ of each dual face using the group multiplication. Indeed, it appears in \eqref{BC action 1} that the Spin(4) element $Q_{f\tau}$ defined by $Q_{f\tau}(q_{f\tau},h_\tau) = (h_\tau q_{f\tau} h_\tau^{-1},q_{f\tau})$ is a source of curvature acting at the level of the dual edge $\tau$. Notice that in the 'time gauge', $h_\tau=\mathrm{id}$, $Q_{f\tau}$ is a spatial rotation, whereas we can anticipate that it will be a boost in the non-geometric sector (see section \ref{non-geom sector}). We thus propose to insert these sources into the holonomy around each face, starting from the action \eqref{BF action} for pure BF theory:
\be \label{BC star action}
S_{\star,BC} = \sum_f \Tr\Bigl(B_f(v_0)\ G_{v_0\tau_1}\,Q_{f\tau_1}\,G_{\tau_1v_1}G_{v_1\tau_2}\,Q_{f\tau_2}\dotsm Q_{f\tau_n}\,G_{\tau_n v_0}\Bigr)
\ee
where $Q_{f\tau}$, naturally defined in the frame of $\tau$, is constrained to be of the form: $Q_{f\tau}(q_{f\tau},h_\tau) = (h_\tau q_{f\tau} h_\tau^{-1},q_{f\tau})$.
\be \begin{split}
S_{\star,BC} = \sum_f \tr\Bigl(b_{-f}(v_0)\ g_{-v_0\tau_1}\, q_{f\tau_1}\, &g_{-\tau_1 v_1}\, g_{-v_1\tau_2}\, q_{f\tau_2}\dotsm\Bigr) \\ + &\tr\Bigl(b_{+f}(v_0)\ g_{+v_0\tau_1}\, h_{\tau_1}q_{f\tau_1}h_{\tau_1}^{-1}\, g_{+\tau_1 v_1}\, g_{+v_1 \tau_2}\, h_{\tau_2} q_{f\tau_2} h_{\tau_2}^{-1}\dotsm\Bigr) \end{split}
\ee
if $v_0,\tau_1,v_1,\tau_2,\dotsc,\tau_n$ are the ordered boundary simplices of $f$ starting at the vertex of reference $v_0$. The variables $L_{fv}$ have thus been eliminated. Notice that in contrast with \eqref{BC action} which is well-defined as soon as the faces (and dual edges) are oriented, the action \eqref{BC star action} also requires a base-point $v_0$ for each face (and thus an ordering). This action {\it a priori} depends on the choice of the base-points. Indeed, if a bivector is written in another frame using \eqref{def bivectors}, the change cannot be compensated via \eqref{transport hol} as it is the case in pure BF theory, because of the presence of the external sources $q_{f\tau}$ (external here refers to the fact they do not come from a discretization of the connection). For this to work, we would need to define rules for parallel transport involving $Q_{f\tau}$. However, it will be clear that the partition function does not depend on the base-points.

Due to the linearity of \eqref{BC star action} in the variables $B_f(v_0)$, the partition functions projects the group elements coupled to them to the identity:
\begin{align}
Z_{BC} &= \int \prod_{(\tau,v)} dG_{\tau v} \int \prod_{\tau} dh_\tau \prod_{(f,\tau)} dq_{f\tau} \int \prod_f dB_f(v_0)\ \prod_f e^{i\Tr(B_f(v_0)\,G_{v_0\tau_1}\,Q_{f\tau_1}(q_{f\tau_1},h_{\tau_1})\,G_{\tau_1v_1}\dotsm)} \\
 &= \int \prod_{(\tau,v)} dG_{\tau v} \int \prod_{\tau} dh_\tau \prod_{(f,\tau)} dq_{f\tau}\ \prod_f \delta\bigl(g_{-v_0\tau_1}\,q_{f\tau_1}\,g_{-\tau_1 v_1}\,g_{-v_1\tau_2}\, q_{f\tau_2}\dotsm\bigr) \delta\bigl(g_{+v_0\tau_1}\,h_{\tau_1}q_{f\tau_1}h_{\tau_1}^{-1}\,g_{+\tau_1v_1}\dotsm\bigr)
\end{align}
It is convenient at this stage to make some changes of variables. First, we redefine the variables $q_{f\tau}$ to $g_{-s(\tau)\tau}\,q_{f\tau}\,g_{-\tau t(\tau)}$, so that the delta functions on the anti-self-dual sector simply read: $\delta(q_{f\tau_1}\,q_{f\tau_2}\dotsm)$. The delta functions on the self-dual sector then contains the combinations: $h_\tau(s(\tau)) = g_{+s(\tau) \tau}\ h_\tau\ g_{-s(\tau) \tau}^{-1}$ and $h_\tau(t(\tau)) = g_{+\tau t(\tau)}^{-1}\ h_\tau\ g_{-\tau t(\tau)}$ which are the result of parallelly transporting $h_\tau$ to the source vertex $s(\tau)$ of the dual edge $\tau$ and to its target vertex $t(\tau)$. Since the group elements $g_{+v\tau}$ only enters the action this way, we use the Haar measure to absorb $h_\tau\ g_{-s(\tau) \tau}^{-1}$ and $h_\tau\ g_{-\tau t(\tau)}$ into $g_{+s(\tau)\tau}$ and $g_{+\tau t(\tau)}$ respectively. We thus define:
\be \label{decoupling normals}
h_{\tau,s(\tau)} = g_{+s(\tau)\tau}\ h_\tau\ g_{-s(\tau)\tau}^{-1} \qquad \text{and}\qquad h_{\tau,t(\tau)} = g_{+\tau t(\tau)}^{-1}\ h_\tau\ g_{-\tau t(\tau)}
\ee
These $\SU(2)$ group elements now define two vectors $N_{\tau,t(\tau)}$ and $N_{\tau,s(\tau)}$, both interpreted as orthogonal to $\tau$, but seen as independent from each other, according to the frame of each 4-simplex sharing it. This change of variables explains that from the effective point of view, upon integrating the holonomy degrees of freedom, the BC model deals with seemingly isolated 4-simplices. It is equivalent to say that each $h_\tau$ and $g_{-\tau v}$ can be gauge-fixed to the identity.

The partition function now takes the form:
\begin{align}
Z_{BC} &= \int \prod_{(\tau,v)} dh_{\tau,v} \int \prod_{(f,\tau)} dq_{f\tau}\ \prod_f \delta\bigl(q_{f\tau_1}\, q_{f\tau_2}\dotsm q_{f\tau_n}\bigr)\ \delta\bigl(h_{\tau_1,v_0}\,q_{f\tau_1}\,h_{\tau_1,v_1}^{-1}\,h_{\tau_2,v_1}\dotsm q_{f\tau_n}\,h_{\tau_n,v_0}^{-1}\bigr) \\
&= \sum_{\{j_{\pm f}\}} \int \prod_{(\tau,v)} dh_{\tau,v} \int \prod_{(f,\tau)} dq_{f\tau}\ \prod_f d_{j_{+f}}d_{j_{-f}}\,\chi_{j_{-f}}\bigl(q_{f\tau_1}\, q_{f\tau_2}\dotsm q_{f\tau_n}\bigr)\,\chi_{j_{+f}}\bigl(h_{\tau_1,v_0}\,q_{f\tau_1}\,h_{\tau_1,v_1}^{-1}\,h_{\tau_2,v_1}\dotsm q_{f\tau_n}\,h_{\tau_n,v_0}^{-1}\bigr) \label{BC last step}
\end{align}
In the second line, we have expanded the delta functions over $\SO(3)$ into characters, with $j_{+f},j_{-f}\in\mathbbm{N}$. Each $q_{f\tau}$ appears twice, so that the othogonality relation \eqref{orthogonality} perfectly works to integrate them. This selects the simple representations, $j_{-f}=j_{+f}$, and recombines into characters the insertions $h_{\tau_i,v_i}^{-1} h_{\tau_{i+1},v_i}$ between the $q_{f\tau}$s. To see it, notice that the formula \eqref{orthogonality} is simple enough so that we can keep track of the indices. Indeed, the right $\SU(2)$ index of $q_{f\tau_1}$ in the anti-self-dual part is, due to \eqref{orthogonality}, the same as that of $q_{f\tau_1}$ in the self-dual part, which is contracted with $h_{\tau_1,v_1}^{-1}h_{\tau_2,v_1}$. Then, the r.h.s. of the latter is contracted with the l.h.s. of $q_{f\tau_2}$ which is also contracted, via \eqref{orthogonality}, in the anti-self-dual part, with the right index of $q_{f\tau_1}$. This thus forms the character $\chi_{j_f}(h_{\tau_1,v_1}^{-1}h_{\tau_2,v_1})$. Let us perform explicitly the computation for a triangular face. We are interested in the following quantity:
\begin{align}
&\int \prod_{i=1}^3 dq_i\ \chi_{j_-}\bigl(q_1\, q_2\, q_3\bigr)\,\chi_{j_+}\bigl(q_1\ \tl{h}_{12}^{-1}\,h_{12}\ q_2\ \tl{h}_{23}^{-1}\,h_{23}\ q_3\ \tl{h}_{31}^{-1}\,h_{31}\bigr) \\
& = \sum_{\substack{a_{1,2,3} \\ \alpha_{1,2,3},\beta_{1,2,3}}} \int \prod_{i=1}^3 dq_i\ D^{(j_-)\star}_{a_3 a_1}(q_1)\,D^{(j_-)\star}_{a_1 a_2}(q_2)\,D^{(j_-)\star}_{a_2 a_3}(q_3)\,D^{(j_+)}_{\alpha_3 \beta_1}(q_1)\,D^{(j_+)}_{\alpha_1 \beta_2}(q_2)\,D^{(j_+)}_{\alpha_2 \beta_3}(q_3)\,\prod_{i=1}^3 D^{(j_+)}_{\beta_i \alpha_i}(\tl{h}_{i\,i+1}^{-1}\,h_{i\,i+1}) \\
&= \f{\delta_{j_+,j_-}}{d_{j_+}^3} \sum_{\substack{a_{1,2,3} \\ \alpha_{1,2,3},\beta_{1,2,3}}} \prod_i \delta_{a_i,\alpha_i}\delta_{a_i\beta_i}\ D^{(j_+)}_{\beta_i \alpha_i}(\tl{h}_{i\,i+1}^{-1}\,h_{i\,i+1}) \\
&= \f{\delta_{j_+,j_-}}{d_{j_+}^3}\ \chi_{j_+}(\tl{h}_{12}^{-1}\,h_{12})\,\chi_{j_+}(\tl{h}_{23}^{-1}\,h_{23})\,\chi_{j_+}(\tl{h}_{31}^{-1}\,h_{31})
\end{align}
Thus, \eqref{BC last step} precisely leads us to the integral representation of the 10j-symbol:
\begin{align}
Z_{BC} &= \sum_{\{j_f\}} \prod_f d_{j_f}^2 \prod_\tau \f{1}{\prod_{f\subset\pp\tau}d_{j_f}}\ \prod_v \Big[\int \prod_{(\tau,v)} dh_{\tau,v}\ \prod_{f\subset \pp v} \chi_{j_f}\big(h_{u(v),v}\ h^{-1}_{d(v),v}\big)\Big] \label{BC}\\
 &=  \sum_{\{j_f\}} \prod_f d_{j_f}^2 \prod_\tau \f{1}{\prod_{f\subset\pp\tau}d_{j_f}}\ \prod_v 10j\big(j_f\big)
\end{align}
where $u(v)$ and $d(v)$ are the two tetrahedra sharing $f$ in the 4-simplex $v$.

The action \eqref{BC star action} thus provides the BC model with a very natural action of the BF type. The group elements $h_{\tau,v}$ entering the integral representation of the 10j-symbol represent the unit vectors $N_{\tau,v}$ orthogonal to the faces of each tetrahedron $\tau$ when it is seen in the frame of the 4-simplex $v$. The amplitude then compares the way $N^{(0)}$ is mapped into $N_{\tau,v}$ for adjacent tetrahedra within each 4-simplex. Because the change of variables \eqref{decoupling normals} is made on the group elements $g_{+v\tau}$, the variables $h_{\tau,v}$ can also be seen as $\SU(2)$ holonomies between $\tau$ and $v$, with $h_\tau$ and $g_{-v,\tau}$ being gauge-fixed to the identity, that is with flat anti-self-dual transport and orthogonal directions $N_\tau=N^{(0)}$ to tetrahedra.

The obtained amplitude for tetrahedra has already been used in \cite{group integral techniques} and proposed in \cite{gluing oriti} as an alternative to those of \cite{perez-rovelli} and \cite{depietri-etal}. In particular, it does not contain the so-called 'eye diagram'. It corresponds to a trivial gluing between 4-simplices in the language of \cite{gluing oriti}. Notice however that our derivation uses from the beginning a triangulation and not isolated 4-simplices. The decorrelation between the points of view of 4-simplices sharing a tetrahedron comes from the integration over the holonomy degrees of freedom. We also emphasize that the method used to impose the constraints at the level of the action, for \eqref{BC action 1} and \eqref{BC star action}, realizes a natural implementation from the spin foam point of view, which can be seen as a weak implementation since it respects the group structure of $\SU(2)$ entering the BF action. It is shown in appendix \ref{naive BC} that, from the point of view of the action \eqref{BC action}, it corresponds to introducing the constraints into the path integral with the measure \eqref{bessel measure}, which is not surprising in the view of section \ref{sec simple rep}. As for insertions of observables, it is clear that the actions \eqref{BC action 1} and \eqref{BC star action} produce the same results for insertions of holonomies. As for $B$-dependent observables, one has to be careful since both actions are defined with different bivectors.

\subsection{Equations of motion for the BC action} \label{eom BC}

Having shown that the BC model can be derived by imposing the constraints on disjoint tetrahedra, one can still clarify the situation described by the model, and in particular the gluing process, by looking at the equations of motion of the discrete action. Instead of \eqref{BC star action} which requires an explicit ordering of the dual edges, let us consider the action \eqref{BC action 1} which treats tetrahedra more symmetrically. We obviously have:
\be \label{trivial eom}
q_{f\tau}=g_{-f\tau}^{-1} \qquad \text{and} \qquad g_{+f\tau}\,h_\tau\, g_{-f\tau}^{-1}\,h_\tau^{-1}=\mathrm{id}
\ee
which still hold at the quantum level since \eqref{BC action 1} is linear in the bivectors. These can be thought of as characterizing the model, exactly as $\delta(G_f)$ defines the BF spin foam model and $\delta(g_{+f}h_fg_{-f}^{-1}h_f^{-1})$ restricts to simple bivectors. If we interpret $G_{f\tau}$ as parallel transport around $f$ seen in $\tau$, the right equation of \eqref{trivial eom} means that $N_\tau$ is left invariant by parallel transport. In the time gauge, it says that $G_{f\tau}$ is simply a spatial rotation. Notice that in the continuum, the equation of motion w.r.t. the field $B$ still involves $B$ (since the constraints are quadratic in $B$). However, the analogous equation is here independent of the bivectors.

To extremize the action w.r.t. group elements such as $q_{f\tau}$, we use right invariant vector fields given by $\delta q_{f\tau}\,q_{f\tau}^{-1}\in\su(2)$. Extremizing the action w.r.t. the multipliers $q_{f\tau}$ (and using \eqref{trivial eom}) naturally gives the simplicity constraints \eqref{cross-simplicity BC}. The equations of motion for $G_{\tau t(\tau)}$ and $G_{s(\tau)\tau}$ are respectively:
\be \label{closure}
\sum_{f\subset \pp\tau} B_{f\tau} = 0\qquad \text{and}\qquad \sum_{f\subset \pp\tau} G_{f\tau}\, B_{f\tau}\, G_{f\tau}^{-1} = 0
\ee
These relations are naturally consistent with \eqref{cross-simplicity BC} and \eqref{trivial eom}. In agreement with the remarks below \eqref{decoupling normals}, these relations for the self-dual sector only, together with \eqref{cross-simplicity BC} and \eqref{trivial eom}, imply those for the anti-self-dual sector. The first equation of \eqref{closure} is the so-called closure relation and can be seen as a discretization of the continuous equation $d_AB=0$, integrated over a tetrahedron of the triangulation. For pure BF theory, the second equation of \eqref{closure} is equivalent to the closure relation since $G_{f\tau}=\mathrm{id}$. However, due to the sources $q_{f\tau}$, this is not the case anymore and it means that the closure relation is maintained when 'parallelly transporting' each of the four bivectors around the corresponding triangle with $G_{f\tau}$.

Equations \eqref{cross-simplicity BC} and \eqref{closure} ensure metricity for each tetrahedron independently. Let us thus study the regluing based on the variables $L_{fv}$. The standard spin foam model for BF theory is often built using some division of the dual faces into plaquettes $p$, which allows the assignment of independent variables, like bivectors, to each piece $p$ of every face. Such a division corresponds to breaking up the original triangulation into a disjoint union of simplices. For instance, in the specific situation here studied, the triangulation has been initially broken up into a disjoint union of tetrahedra, so that a copy of a triangle is specified by a tetrahedron having it in its boundary. The wedge variables are similar, but constructed with pairs $(f,v)$ (see \cite{BF action principle} for instance). Then, we can assign independent bivectors $B_{f\tau}$ to the pairs $p=(f,\tau)$, which are related by group variables, here the elements $L_{fv}$. These latter depend on the dual face and live along the boundary shared by a pair of plaquettes belonging to the same face. We emphasize that they should be seen as a tool for gluing simplices rather than a tool for parallel transport between frames. Such a gluing is however a non-trivial operation. {\it A priori}, the two following actions for BF theory are different:
\be
\sum_{p} \Tr\Big(B_p(v) G_p(v)\Big) \neq \sum_f \Tr\Big(B_f(v) G_f(v)\Big)
\ee
The fact that they both lead to the same spin foam model is due to the specificity of BF theory, that is the triviality of the holonomies. For adjacent plaquettes of a face $f$, the bivectors are related by:
\be \label{gluing}
B_{f,\tau_2} = G_{\tau_1 \tau_2}^{-1}\,B_{f,\tau_1}\,G_{\tau_1 \tau_2}
\ee
when the action \eqref{broken up action} is extremized w.r.t. to $L_{fv}$. Such a relation is necessary for consistency between frames and is a discretization of the continuous e.o.m. $d_A B=0$ which states that $B$ is constant up to parallel transport. Moreover, implementing it all along the boundary of $f$, it implies that the holonomy $G_f$ leaves $B_f$ invariant, $G_f\,B_f\,G_f^{-1} = B_f$. The fact that the degrees of freedom of $G_f$ lie in $\U(1)_{b_{+f}}\times\U(1)_{b_{-f}}$ is a basic property of Regge calculus.

We expect this relation and \eqref{gluing} to be modified in the presence of a source for parallel transport. Indeed, the variables $q_{f\tau}$ are responsible for a new relation when extremizing w.r.t. $L_{fv}$:
\be \label{relating bivectors}
B_{f,\tau_2} = Q_{f,\tau_2}^{-1}\, G_{\tau_1 \tau_2}^{-1}\, B_{f,\tau_1}\, G_{\tau_1 \tau_2}\, Q_{f,\tau_2}\quad\text{with}\quad Q_{f\tau} = \big(h_\tau q_{f\tau} h_\tau^{-1},q_{f\tau}\big)
\ee
This is precisely the relation which makes the action \eqref{BC star action} independent of the base-points. This specific relation \eqref{relating bivectors} induces a non-trivial effect on the relation \eqref{coupling tetra} between the normal vectors of neighbouring tetrahedra. Using simplicity under the form \eqref{cross-simplicity BC} together with \eqref{relating bivectors}, one obtains that $h_{\tau_2}(\tau_1) h_{\tau_1}^{-1}(\tau_1) = g_{+\tau_1\tau_2}h_{\tau_2}g_{-\tau_1\tau_2}^{-1}h_{\tau_2}^{-1}$ leaves $b_{+f\tau_1}$ invariant, as expected (or equivalently, the Spin(4) element $H_{\tau_1\tau_2}(\tau_1)=(h_{\tau_2}(\tau_1) h_{\tau_1}^{-1}(\tau_1), h_{\tau_1}^{-1}(\tau_1)h_{\tau_2}(\tau_1))$ preserves the bivector $B_{f\tau_1}$). But transporting this relation in the frame of $\tau_2$ with \eqref{relating bivectors} shows that $h_{\tau_2}(\tau_2) h_{\tau_1}^{-1}(\tau_2)=h_{\tau_2}g_{-\tau_1\tau_2}^{-1}h_{\tau_1}^{-1}g_{+\tau_1\tau_2}$ does not leave $b_{+f,\tau_2}$ invariant, due to insertions of $q_{f\tau}$. We have instead:
\begin{align} \label{corr BC}
&h_{\tau_2}\,q_{f\tau_2}^{-1}\,h_{\tau_1}^{-1}(\tau_2)\ \tl{b}_{+f,\tau_2}\ \bigl(h_{\tau_2}\,q_{f\tau_2}^{-1}\,h_{\tau_1}^{-1}(\tau_2)\bigr)^{-1} =\ b_{+f,\tau_2} \\
&\text{with}\qquad\qquad\qquad \tl{b}_{+f\tau_2} = h_{\tau_2}q_{f\tau_2}h_{\tau_2}^{-1}\ b_{+f\tau_2}\ h_{\tau_2}q_{f\tau_2}^{-1}h_{\tau_2}^{-1}
\end{align}
or equivalently, $(h_{\tau_2}\,q_{f\tau_2}^{-1}\,h_{\tau_1}^{-1}(\tau_2),q_{f\tau_2}^{-1}\,h_{\tau_1}^{-1}\,h_{\tau_2}(\tau_2))$ acting on $Q_{f\tau_2}\,B_{f\tau_2}\,Q_{f\tau_2}^{-1}$ gives $B_{f\tau_2}$.

To conclude this study, it is clear that the action \eqref{BC action 1} (weakly) imposes cross-simplicity on disjoint tetrahedra. The variables $L_{fv}$ have to reglue them together, but precisely because of simplicity, we do not obtain the expected relations when parallelly transporting the bivectors describing the same triangle. The variables $q_{f\tau}$ play indeed the role of holonomies but with a dependence on the face. This in turn alters the correlations between neighbouring tetrahedra. The rotation $h_{\tau_2}h_{\tau_1}^{-1}$ is supposed to preserve $b_{+f}$, since the two normals live in the same plane orthogonal to $\star B_f$. However, this relation is not satisfied in every frame.

\subsection{In the non-geometric sector} \label{non-geom sector}

The explicit implementation of the classical relation \eqref{cross-simplicity BC} into the path integral shows that the BC model is a quantization on isolated tetrahedra of the geometric sector of Plebanski's theory. The non-geometric sector corresponds to:
\be \label{cross-simplicity topo}
b_{-f\tau} = h_\tau^{-1}\ b_{+f\tau}\ h_\tau
\ee
Since $-\tr(bg)=\tr(bg^{-1})$, it is sufficient to replace each $q_{f\tau}$ with its inverse in the self-dual (or equivalently the anti-self-dual) part of the action \eqref{BC action 1} to obtain this sector. Working again on disjoint tetrahedra, we propose:
\be \label{ng action 1}
\tl{S} = \sum_{(f,\tau)} \tr\bigl(b_{-f\tau}\,g_{-f\tau}\,q_{f\tau}\bigr) + \tr\bigl(b_{+f\tau}\,g_{+f\tau}\,h_\tau\, q_{f\tau}^{-1}\,h_\tau^{-1}\bigr)
\ee

\begin{figure} \begin{center}
\includegraphics[width=5cm]{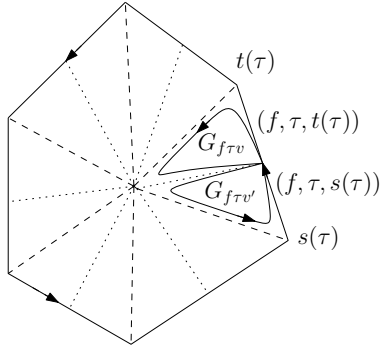} 
\caption{ \label{half-wedges} It represents a dual face, whose edges, denoted $\tau$, are dual to tetrahedra and vertices dual to 4-simplices. The dotted lines delimit the wedges, that are pairs $(f,v)$ and which correspond to the triangles of a triangulation initially broken up into 4-simplices. The dashed lines represent the breaking up of each 4-simplex into disjoint tetrahedra and form the half-wedges, consisting in triplets $(f,\tau,v)$. The standard spin foam model for BF theory can be defined with these half-wedges, with the help of holonomies $G_{\tau v}$ and of 'gluing' variables $L_{f\tau}$ and $L_{fv}$ living on the internal links.}
\end{center}
\end{figure}

However, the model defined in this way turns out to be ill-defined. It indeed cruelly depends on the parity of the number of boundary edges of each dual face. This difficulty can be overcome by imposing cross-simplicity not simply on isolated tetrahedra, but instead for each tetrahedron in each 4-simplex separately. The details of the construction are given in appendix \ref{app topo sector}. A tetrahedron $\tau$ being shared by exactly two 4-simplices, $s(\tau)$ and $t(\tau)$, it means that the existence of the unit normal $N_\tau$ is separately imposed in $s(\tau)$ and $t(\tau)$, with the same rotation $h_\tau$, but with different multipliers $q_{f,\tau,s(\tau)}$ and $q_{f,\tau,t(\tau)}$. The flipping of $q$ in the self-dual part of the action can be recorded in $\tl{Q}_{f\tau v} = (h_\tau\,q_{f\tau v}^{-1}\,h_\tau^{-1},q_{f\tau v})$. This Spin(4) element is a boost in the 'time gauge', while it was a spatial rotation in the geometric sector.
\be
\tl{S} = \sum_{(f,\tau,v)} \Tr\Bigl(B_{f\tau v}\ G_{f\tau v}\,\tl{Q}_{f\tau v}\Bigr)
\ee
The assignment of the variables is illustrated in figure \ref{half-wedges}. The plaquettes $(f,\tau,v)$ will be called half-wedges, since they arise from superimposing the wedge division and that into pairs $(f,\tau)$. Since a tetrahedron is shared by two 4-simplices, the number of copies of a triangle is twice that obtained with the pieces $(f,\tau)$. Thus, with regards to the above-mentioned parity problem of the number of dual edges, this construction selects the case where each face has an even number of edges, since all happens as if tetrahedra were chopped into two pieces. Notice that this framework could have also be used in the geometric sector, leading again to the BC model. In particular, the e.o.m. relating the variables $B_{f\tau v}$ are of the type \eqref{relating bivectors} and the variables $Q_{f\tau v}$ can be naturally composed to form: $Q_{f\tau} = Q_{f\tau s(\tau)}Q_{f\tau t(\tau)}$, which is consistent since the normal to $\tau$ is encoded in the same rotation $h_\tau$ for $s(\tau)$ and $t(\tau)$.

Like in the geometric sector, we can 'stick' tetrahedra thinking of the $\tl{Q}_{f\tau v}$s as insertions of curvature source into the holonomies around the faces at each half-wedge, and replace each $Q_{f\tau}$ in \eqref{BC star action} with $\tl{Q}_{f\tau s(\tau)}\,\tl{Q}_{f\tau t(\tau)}$:
\be \label{topo action}
\tl{S}_\star = \sum_f \Tr\Bigl(B_{f}(v_0)\ G_{v_0 \tau_1}\,\tl{Q}_{f\tau_1 v_0}\,\tl{Q}_{f\tau_1 v_1}\,G_{\tau_1 v_1}\,G_{-v_1 \tau_2}\,\tl{Q}_{f\tau_2 v_1}\dotsm\Bigr)
\ee
This gives, in self-dual/anti-self-dual decomposition:
\be  \begin{split}
\tl{S}_\star = \sum_f &\tr\Big(b_{-f}\ g_{-v_0 \tau_1}\ q_{f\tau_1v_0}\ q_{f\tau_1v_1}\ g_{-\tau_1 v_1}\ g_{-v_1 \tau_2}\ q_{f\tau_2v_1}\dotsm\Big) \\
 &+\tr\Big(b_{+f}\ g_{+v_0 \tau_1}\ h_{\tau_1}\, q_{f\tau_1v_0}^{-1}\, q_{f \tau_1 v_1}^{-1}\, h_{\tau_1}^{-1}\ g_{+\tau_1 v_1}\ g_{+v_1 \tau_2}\ h_{\tau_2}\,q_{f \tau_2 v_1}^{-1}\dotsm\Big)
\end{split} \ee
as an equivalent of \eqref{BC star action} in the non-geometric sector. In the geometric sector, $q_{f\tau v}^{-1}$ is simply replaced with $q_{f\tau v}$ in the self-dual part, so that they can be obviously composed to form the multiplier $q_{f\tau}\equiv q_{f \tau s(\tau)} q_{f \tau t(\tau)}$. However, this is not possible anymore in the non-geometric sector because of the flipping of the multipliers in the self-dual part of the action.

The steps of the computation are similar to those of the derivation of the BC model. First integrate the bivectors:
\be \begin{split}
\tl{Z} = \int \prod_{(\tau,v)} dG_{v\tau} \int \prod_\tau dh_\tau \prod_{(f,\tau,v)} dq_{f\tau v} &\prod_f \delta\bigl(g_{-v_0 \tau_1}\ q_{f\tau_1v_0}\ q_{f\tau_1v_1}\ g_{-\tau_1 v_1}\ g_{-v_1 \tau_2}\ q_{f\tau_2v_1}\dotsm\bigr) \\
&\delta\bigl(g_{+v_0 \tau_1}\ h_{\tau_1}\, q_{f\tau_1v_0}^{-1}\, q_{f\tau_1v_1}^{-1}\, h_{\tau_1}^{-1}\ g_{+\tau_1 v_1}\ g_{+v_1 \tau_2}\ h_{\tau_2}\,q_{f\tau_2v_1}^{-1}\dotsm\bigr)
\end{split} \ee
and expand the resulting expression into characters. Then integrate over the variables $q_{f,\tau,v}$, thus selecting the simple representations. To see more precisely what happens, let us keep track of $\SU(2)$ indices while using \eqref{orthogonality}. The right index of $g_{-v_0\tau_1}$ is contracted with the left index of $q_{f\tau_1 v_0}$. Thanks to \eqref{orthogonality}, the latter is equal to the left index of $q_{f\tau_1 v_1}^{-1}$ in the self-dual sector. Using \eqref{orthogonality} again, this index equals the right index of $q_{f\tau_1v_1}$ in the anti-self-dual sector, which is contracted with $g_{-\tau_1v_1}$, and so on \dots This process forms the holonomy $g_{-f}(v_0)$. In the self-dual sector, the rotations $h_\tau$ disappear and we also end up with $g_{+f}(v_0)$. The remaining integrations are taken to be over the group elements $G_\tau \equiv G_{s(\tau)\tau} G_{\tau t(\tau)}$ allowing for parallel transport between 4-simplices, with $G_f=G_{\tau_1}G_{\tau_2}\cdots$. Taking into account the dimensional factors $d_{j_f}$, we have:
\begin{align}
\tl{Z} &= \int\prod_\tau dG_{\tau}\ \sum_{\{j_f\}}\ \prod_\tau \f{1}{\prod_{f\subset\pp\tau}d_{j_f}^2}\ \prod_f d_{j_f}^2\ \chi_{j_f}\bigl(g_{+f}\bigr)\ \chi_{j_f}\bigl(g_{-f}\bigr) \\
 &= \sum_{\{j_f\}}\sum_{\{(i_{+\tau},i_{-\tau})\}}\ \prod_f d_{j_f}^2\ \prod_\tau \f{1}{\prod_{f\subset\pp\tau}d_{j_f}^2}\ \prod_v 15j_{\mathrm{Spin(4)}}\bigl((j_f,j_f);(i_{+\tau},i_{-\tau})\bigr) \label{topo model}
\end{align}
The model can thus be straightforwardly interpreted as a spin foam model for BF theory restricted to the simple representations of Spin(4). However, in comparison with the model obtained in section \ref{sec simple rep} with the diagonal simplicity constraints, the face and tetrahedron amplitudes are different and are the direct effects of the constraints\footnotemark \eqref{cross-simplicity topo}.

\footnotetext{Let us briefly present a model based on $\SU(2)$ BF theory in which face or tetrahedron amplitudes are affected by adding some information. Let $\phi_f$ and $\psi_f$ be in $\SU(2)$, and add their commutator $\phi_f\psi_f\phi_f^{-1}\psi_f^{-1}$ to the usual discrete action:
\be
S = \sum_f \tr\bigl(b_f\,g_f\,\phi_f\,\psi_f\,\phi_f^{-1}\,\psi_f^{-1}\bigr)
\ee
The interpretation of the model is as follows. The e.o.m. are:
\be
g_f\,\phi_f\,\psi_f\,\phi_f^{-1}\,\psi_f^{-1} = \mathrm{id}\qquad \phi_f\,b_f\,\phi_f^{-1} = \psi_f\,b_f\,\psi_f = b_f
\ee
together with the closure relation. Thus $\phi_f$ and $\psi_f$ give a source of curvature which is classically constrained to live in the $\U(1)$ subgroup preserving $b_f$. In this case, $\phi_f$ and $\psi_f$ commute, so that $g_f=\mathrm{id}$. However, at the quantum level, $g_f$ is not constrained to be the identity. Although the partition function does not project on flat connection anymore, the vertex of the resulting spin foam model is still the 15j-symbol:
\be
Z = \sum_{\{j_f\}} \prod_f \f{1}{d_{j_f}} \prod_v 15j(j_f)
\ee
Only the face weight, $d_{j_f}^{-1}$ instead of $d_{j_f}$, takes into account the modification of the action. In this regard, notice that it is well-known (since the works of Ponzano and Regge) that the face weight $d_{j_f}$ is required for the model to be triangulation independent.
}

\subsection{About the Immirzi parameter} \label{immirzi}

The Immirzi parameter $\gamma$ represents a one-parameter ambiguity in classical pure gravity. It can be included into the usual Palatini-Cartan action by using the Hodge dual $\star$ of $\so(4)$: add to the curvature $F$ the quantity $\star F$ with coefficient $\gamma$:
\be \label{holst}
S_{\mathrm{Holst}} = \int \eps_{IJKL}\ e^I\wedge e^J\wedge \l(F^{KL}+2\gamma (\star F)^{KL}\r)
\ee
This is the so-called Holst action \cite{holst}. The equation of motion for the new term in this action is simply: $d_A(e\wedge e)=0$, which clearly does not modify the original equations. Notice that it is equivalent to leave $F$ unchanged and proceed to $e\wedge e\rightarrow e\wedge e+\gamma\star(e\wedge e)$ instead. In Plebanski's formulation, the field $B$ is then similarly transformed, but the simplicity constraints also need to be changed. We thus prefer to use the version \eqref{holst} with $B=\star (e\wedge e)$ together with the usual simplicity constraints.

Let us define for $g = \cos\theta\,\mathrm{id}+i\sin\theta\,\hat{n}\cdot\vec{\sigma}\ \in\SU(2)$ and $\alpha\in\mathbb{R}$, the $\SU(2)$ element $g^\alpha$ whose class angle is $\alpha\theta$ and direction $\hat{n}$:
\be \label{imm effect}
g^\alpha := \cos\l(\alpha\theta\r)\,\mathrm{id} + i\sin\l(\alpha\theta\r)\,\hat{n}\cdot\vec{\sigma}
\ee
For integral $\alpha$, this is obviously the group multiplication. The BF action with an Immirzi term reads:
\be \label{bf gamma}
S_\gamma = \int \l(1+\gamma\r) \tr\l( b_+\wedge F_+\r)\ +\ \l(1-\gamma\r) \tr\l( b_-\wedge F_-\r)
\ee
and is clearly equivalent to pure BF theory, but it will not be the case anymore in the presence of the simplicity constraints. The issue is now to introduce $\gamma$ at the discrete level. Consider the $\su(2)$ element $g_{\pm f}$ standing for the discrete curvature, that is the holonomy of the connection along the boundary of the dual face $f$. In a given chart, consider that the edges of $f$ are of order $\varepsilon$. When $\varepsilon$ goes to zero, we have: $g_{\pm f}\approx 1+\varepsilon^2 F_{\pm |f}$, where $F_{\pm |f}$ is the component of $F_\pm$ along the face $f$. Thus, defining $\gamma_\pm = 1\pm\gamma$, it is clear that: $g_{\pm f}^{\gamma_\pm}\approx 1+\varepsilon^2 \gamma_\pm F_{\pm |f}$, and a natural choice to reproduce the continuous action \eqref{bf gamma} is:
\be
S_\gamma = \sum_f \tr\l(b_{+f}\, g_{+f}^{\gamma_+}\r)\ +\ \tr\l(b_{-f}\, g_{-f}^{\gamma_-}\r)
\ee

Let us study the imposition of the diagonal simplicity constraints into this framework, to compare the result with the usual ansatz for the introduction of the Immirzi parameter into spin foams \cite{FKLS,epr imm}. Diagonal simplicity still expresses the fact that the bivectors are simple, and is solved by relating $b_{+f}$ to $b_{-f}$ through a rotation $h_f$ which represent the choice of an vector $N_f$ orthogonal to $\star B_f$. At the classical level, the Immirzi parameter modifies the relation between the self-dual and anti-self-dual parts of the curvature. We have indeed:
\be \label{diag simplicity imm}
b_{-f} = -h_f^{-1}\ b_{+f}\ h_f \qquad\Rightarrow\qquad g_{+f}^{\gamma_+} = h_f\ g_{-f}^{\gamma_-}\ h_f^{-1}
\ee
which means that the vector $N_f$ is not preserved anymore by the Spin(4) element $G_f=(g_{+f},g_{-f})$, as it is the case in \eqref{physical step simple rep}, but by the element $(g_{+f}^{\gamma_+},g_{-f}^{\gamma_-})$. Notice that the quantities $\gamma_\pm$ are not relevant in themselves, but rather their ratio is. Indeed, \eqref{diag simplicity imm} is equivalent to: $g_{+f} = h_f\ g_{-f}^{\gamma_-/\gamma_+}\ h_f^{-1}$, saying that $g_{+f}$ lies in the conjugacy class of $g_{-f}^{\gamma_-/\gamma_+}$. We can then easily read the limits $\gamma\rightarrow \pm\infty$:
\be
g_{+f} = h_f\ g_{-f}^{-1}\ h_f^{-1}
\ee
which indeed corresponds to the non-geometric sector. For a general $\gamma$, since the action and the constraint are linear in the bivectors, \eqref{diag simplicity imm} still holds, avoiding some measure factors like in section \ref{sec simple rep}:
\begin{align}
&\int \prod_{(t,v)} dG_{tv}\ \prod_f \int dh_f\ \delta\l( g_{+f}\ h_f\ g_{-f}^{\gamma_-/\gamma_+}\ h_f^{-1}\r) \\
&= \sum_{\{j_f\}} \int \prod_{(t,v)} dG_{tv}\ \prod_f \chi_{j_f}\bigl(g_{+f}\bigr)\ \chi_{j_f}\l(g_{-f}^{\gamma_-/\gamma_+}\r) \label{Z diag imm}
\end{align}
The resulting spin foam model is however ill-defined for a generic $\gamma$: $g_{-f}$ is a product of group elements $g_{-tv}$ and we do not know how to apply the operation \eqref{imm effect} on each subfactor. We thus ask for a quantization of $\gamma$ such that $n_\gamma=\gamma_-/\gamma_+=(1-\gamma)/(1+\gamma)$ is an integer. The usual ansatz for introducing the Immirzi parameter is based on the canonical analysis of the constraints, in which the bivectors are quantized as the generators of $\so(4)$. This leads to the following sum:
\be \label{imm ansatz}
\sum_{\{j_f\}} \int \prod_{(t,v)} dG_{tv}\ \prod_f \chi_{j_f}\bigl(g_{+f}\bigr)\ \chi_{n_\gamma j_f}\bigl(g_{-f}\bigr)
\ee
with the same quantization of $\gamma$. If the dimension $d_j=2j+1$, instead of the spin, is scaled by $n_\gamma$ in \eqref{imm ansatz}, the difference with \eqref{Z diag imm} can be seen to be a measure ambiguity, and both are easily related. Indeed:
\be
\chi_{k_{j,n}}(g)=\f{\sin d_j n\theta}{\sin \theta}= \chi_{(n-1)/2}(g)\ \chi_j(g^n)
\ee
 for $k_{j,n} = nj+(n-1)/2$ chosen such that: $d_{k_{j,n}} = nd_j$. We thus propose to include the Immirzi parameter with the following action:
\be
\sum_{f} \tr\bigl( b_{+f}\ g_{+f}\, h_f\, q_f\, h^{-1}_f\bigr)\ +\ \tr\bigl( b_{-f}\ g_{-f}^{n_\gamma}\, q_f\bigr)
\ee
where $q_f\in\SU(2)$ is a Lagrange multiplier. Then, we can think of \eqref{imm ansatz} starting from this action with the following weight for $q_f$: $\chi_{(n_\gamma-1)/2}(q^{1/n_\gamma})$, determined by the Immirzi parameter. To deal with cross-simplicity, the basic idea is thus to replace $g_{-f\tau}^{-1}$ in \eqref{BC action 1} with $g_{-f\tau}^{-n_\gamma}$. The computations are quite cumbersome (in particular, the number of appearance of $l_{-fv}$ depends on $\gamma$) and the result is not enlightening, so we do not reproduce them.

\section{Some generalisations of the BC model} \label{sec BC measure}

We present in this section two types of generalisations of the BC model. The first type is a natural extension due to choosing a non trivial measure for the Lagrange multipliers $q_{f\tau}$. It does not change the physical setting of the model, but gives a family of models in a simple way. Furthermore, the vertex amplitude of these models naturally exhibits the fusion coefficients of the new spin foam models \cite{EPR, consistently solving, FKLS}. The second generalisation consists in weakening the imposition of the constraints. The new spin foam models can be interpreted in this framework.

\subsection{The fusion coefficients} \label{sec fusion coeff}

\begin{figure} \begin{center}
\includegraphics[width=6cm]{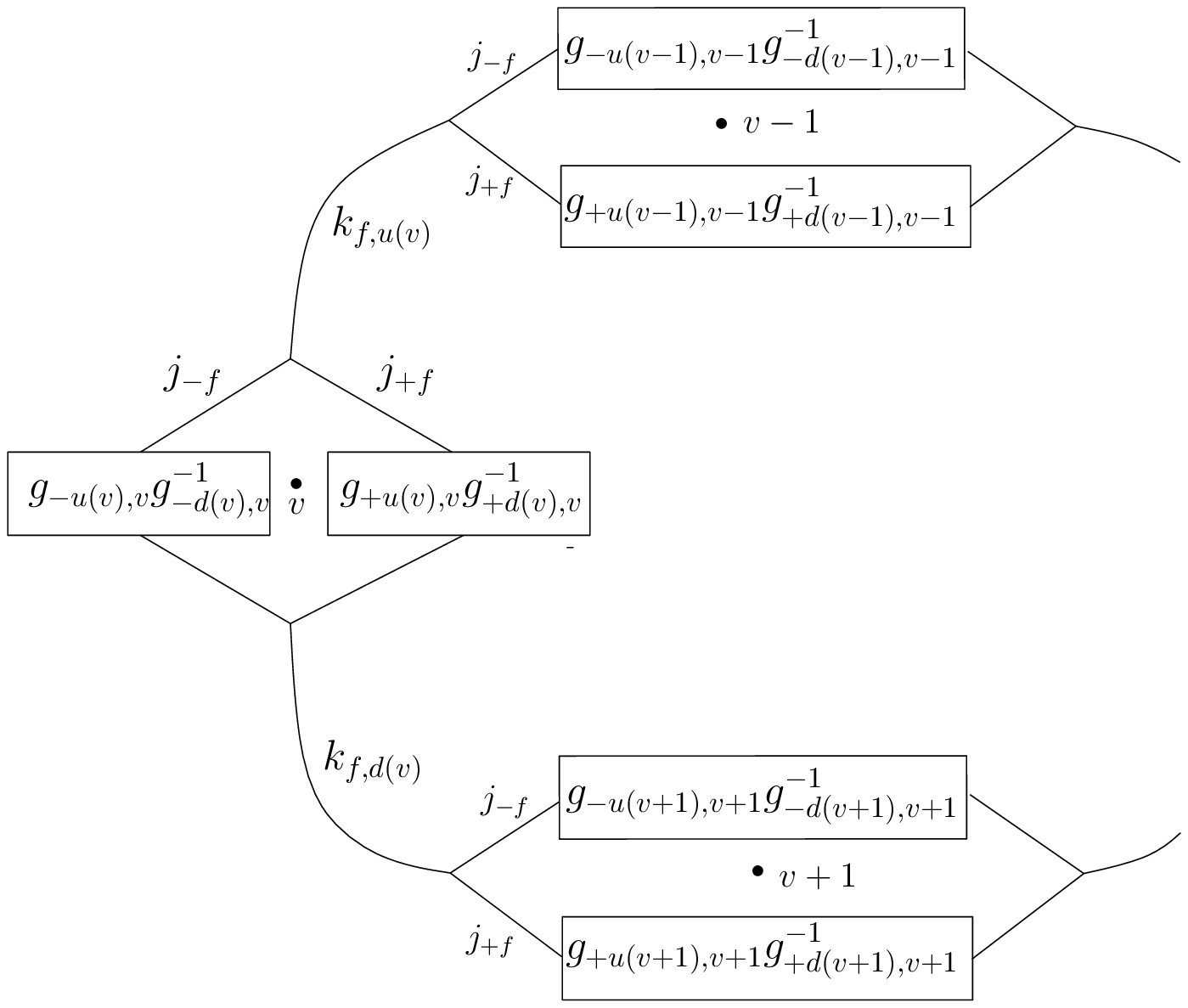}\qquad\qquad\includegraphics[width=10cm]{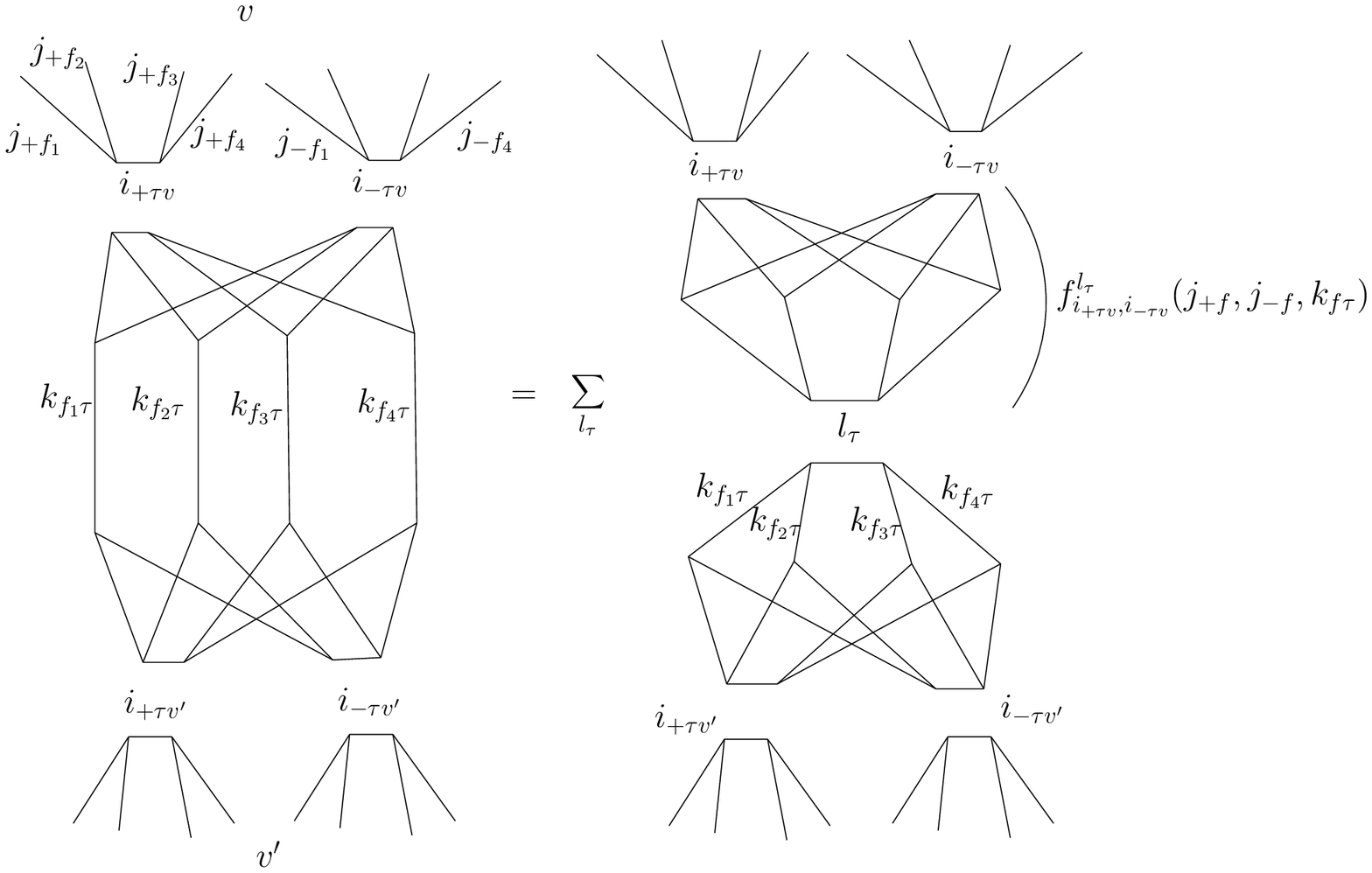}
\caption{ \label{general vertex} The left picture depicts the sequence of couplings for a given face. The links carrying the measure representations $k_{f,\tau}$ are the boundary edges of $f$, dual to tetrahedra. It is a very natural generalisation of the BC model in which all $k_{f,\tau}$ are taken to be zero. The right picture represents the vertex of the model. The faces (the links in the picture) are labelled with irreducible representations $(j_{+f},j_{-f})$ of Spin(4). These are intertwined at each tetrahedron (node in the picture) with the representations $k_{f,\tau}$ which are shared by others vertices.}
\end{center} \end{figure}

We have seen that the action \eqref{BC star action} implements the constraints naturally with regards to the $\SU(2)$ structure and gives a straightforward derivation of the BC model, with a trivial weight for the variables $q_{f\tau}$. Let us still consider the action \eqref{BC star action}, which imposes the existence of normal vectors $N_\tau$ independently for each tetrahedron, and introduce a typical gauge invariant weight $\mu(q_{f\tau})=\chi_{k_{f\tau}}(q_{f\tau})$, where the spins $(k_{f\tau})$ are initially given. Omitting the integrals over the rotations $h_\tau$ and over the holonomies $G_{v\tau}$, the partition function is:
\be
Z_{\{k_{f,\tau}\}} = \int \prod_{(f,\tau)} dq_{f\tau} \chi_{k_{f,\tau}}\big(q_{f\tau}\big) \prod_f \delta\Bigl(g_{-\tau_n\tau_1} q_{f\tau_1} g_{-\tau_1\tau_2}\dotsm q_{f\tau_n}\Bigr)\ \delta\Bigl(h_{\tau_n}^{-1}g_{+\tau_n\tau_1} h_{\tau_1}q_{f\tau_1}h_{\tau_1}^{-1} g_{+\tau_1\tau_2}h_{\tau_2}\dotsm q_{f\tau_n}\Bigr)
\ee
if $v_0,\tau_1,v_1,\tau_2,\cdots,\tau_n$ is the sequence of boundary simplices around $f$. The element of parallel transport $G_{\tau_i \tau_{i+1}}$ is naturally $G_{\tau_i \tau_{i+1}} = G_{\tau_i v_i} G_{v_i \tau_{i+1}}$. Expanding the delta functions into characters assigns representations $j_{+f}$ and $j_{-f}$ to each face. The elements $q_{f\tau}$ now appears thrice. Thus, we no longer have $j_{+f} = j_{-f}$, but instead these representations are intertwined at each tetrahedron with $k_{f\tau}$ according to \eqref{int g3}.
\be \label{first generalisation} \begin{split}
Z_{\{k_{f,\tau}\}} = \sum_{\{(j_{+f},j_{-f})\}} \prod_f d_{j_{+f}} d_{j_{-f}} \prod_{v\in\pp f} &\begin{pmatrix} j_{-f} & j_{+f} & k_{f\tau} \\ a_{f,\tau} & \alpha_{f,\tau} & A_{f,\tau} \end{pmatrix} \begin{pmatrix} j_{-f} & j_{+f} & k_{f,\tau'} \\ b_{f,\tau'} & \beta_{f,\tau'} & A_{f,\tau'} \end{pmatrix} \\
 &\times D^{(j_{-f})}_{a_{f,\tau} b_{f,\tau'}}\Bigl(g_{-\tau v}\,g_{-\tau' v}^{-1}\Bigr) D^{(j_{+f})}_{\alpha_{f,\tau} \beta_{f,\tau'}}\Bigl(g_{+\tau v}\ g_{+\tau' v}^{-1}\Bigr)
\end{split}
\ee
in which the rotations $h_\tau$ have been absorbed on the right of the group elements $g_{+v\tau}$ (or equivalently gauge-fixed to the identity). $\tau$ and $\tau'$ denote the two tetrahedra sharing $f$ in $v$. The sequence of couplings for a face is described in figure \ref{general vertex}.

The typical vertex is easily derived. The integrals over the group elements $g_{\pm \tau v}$ give rise to 4-valent intertwiners $i_{\pm \tau,v}$, between the four representations $j_{\pm f}$ labelling the faces of $\tau$. The $\SU(2)$ indices carried by $v$ are contracted to form a 15j-symbol for Spin(4): $15j_{\mathrm{Spin}(4)}((j_{+f},j_{-f});(i_{+\tau v},i_{-\tau v}))$. On the other hand, the tetrahedron amplitude, depicted in figure \ref{general vertex}, is due to contracting the indices living at each $\tau$. The links carrying the representations $k_{f\tau}$, coming from the intertwining of $j_{+f}$ and $j_{-f}$, relate the 4-simplices $v$ and $v'$ which share $\tau$. Remarkably, this result can be re-expressed in terms of the fusion coefficients $f_{i_+,i_-}^l$ used in the spin foam model of \cite{FKLS}. Indeed, one only needs to insert the identity on the space of intertwiners $j_{f_1}\otimes j_{f_2}\otimes j_{f_3}\otimes j_{f_4}\rightarrow \C$ as a sum over a complete orthogonal basis consisting of 4-valent intertwiners labelled by the internal representation $l_\tau$, as in figure \ref{general vertex}. The partition function then reads:
\be
Z_{\{k_{f,\tau}\}} = \sum_{\{(j_{+f},j_{-f})\}} \prod_f d_{j_{+f}} d_{j_{-f}} \sum_{\{l_\tau\}}\ \prod_v A_v
\ee
with
\be \label{new vertex}
A_v = \sum_{i_{+\tau v},i_{-\tau v}} 15j_{\mathrm{Spin}(4)}\bigl((j_{+f},j_{-f});(i_{+\tau v},i_{-\tau v})\bigr)\ f_{i_{+\tau v},i_{-\tau v}}^{l_\tau}\l(j_{+f},j_{-f},k_{f\tau}\r)
\ee

These models thus form a very natural generalisation of the BC model, involving the same basic idea to implement the constraints. At first sight, these models seem to be close to the unflipped model of \cite{FKLS}. The differences are the following. One should first impose $j_{-f}=j_{+f}$. Second, having obtain a vertex similar to that of \cite{FKLS}, the main point is how the amplitudes $Z_{\{k_{f\tau}\}}$ should be summed. The physical meaning of the representations $k_{f\tau}$ is unclear and a general measure is a complete expansion into characters $\chi_{k_{f\tau}}$. For instance, the vertex amplitude $A_v$ is weighted in \cite{FKLS} with the Clebsch-Gordan coefficients $C^{j_f}_{j_f}{}^{\phantom{-}j_f}_{-j_f}{}^{k_{f\tau}}_0$. However, such a coupling between the face representations and those coming from the expansion of the measure is here not accessible.

The amplitudes \eqref{new vertex} obtained in this section show that the fusion coefficients $f_{i_+,i_-}^l$ naturally arise. Indeed, the simplicity constraints explicitly relate the variables of the self-dual sector to those of the anti-self-dual sector, the Spin(4) covariance being preserved by $h_\tau$. This is achieved in the functional integral through the Lagrange multipliers $q_{f\tau}$. From this point of view, the BC model does not show up these coefficients because it is the simplest model in this class, defined by the weight 1 for each $q_{f\tau}$. When choosing all $k_{f\tau}$ to be zero, the 3jm-symbols of \eqref{first generalisation} become: $\begin{pmatrix} j_- & j_+ & 0 \\ a & \alpha & 0\end{pmatrix} = \f{(-1)^{j_- -a}}{\sqrt{d_{j_+}}}\delta_{j_-, j_+} \delta_{a,-\alpha}$, so that the intertwiners $i_{\pm\tau v}$ can thus be summed to form the well-known BC intertwiner $i_{BC}$.

\subsection{Weakening the constraints}

A second proposition to look at the BC model with a more general perspective is to weaken the imposition of the constraints. This is indeed the main idea from which the new vertices of \cite{consistently solving, EPR} have originated, and which have been proposed to remedy the drawbacks of the BC model. The main approach makes use of the Hamiltonian analysis, first to quantize the bivectors, then to argue that the cross-simplicity constraints, seen as second class constraints in this framework, have to be weakly imposed.

We have already argued that the physical framework of the new vertices is that of the BC model. We thus still impose the constraints tetrahedron by tetrahedron, and search for implementing the constraints in a weaker sense. For this purpose, we use, instead of the variables $q_{f\tau}$, different multipliers for the self-dual and the anti-self-dual parts of the action \eqref{BC star action}, $q_{+f\tau}$ and $q_{-f\tau}$, which are coupled through  gauge invariant functions $\rho(q_{+f\tau},q_{-f\tau})$. This corresponds to the following partition function:
\begin{align}
Z &= \int \prod_{(f,\tau)} dq_{+f\tau} dq_{-f\tau}\ \rho(q_{+f\tau},q_{-f\tau}) \int \prod_f dB_f\ e^{i\tr\big(b_{-f}(v_0) g_{-v_0\tau_1} q_{-f,\tau_1} g_{-\tau_1 v_1}\cdots\big) + i\tr\big(b_{+f}(v_0) g_{+v_0\tau_1} h_{\tau_1} q_{+f,\tau_1} h_{\tau_1}^{-1} g_{+\tau_1 v_1}\cdots\big)} \\
 &= \int \prod_{(f,\tau)} dq_{+f\tau} dq_{-f\tau}\ \rho(q_{+f\tau},q_{-f\tau}) \prod_f \delta\Big(g_{-v_0\tau_1} q_{-f,\tau_1} g_{-\tau_1 v_1}\cdots\Big)\ \delta\Big(g_{+v_0\tau_1} h_{\tau_1} q_{+f,\tau_1} h_{\tau_1}^{-1} g_{+\tau_1 v_1}\cdots\Big)
\end{align}
again omitting the integrals over $h_\tau$ and $G_{v\tau}$. The gauge invariance of $\rho$ reads $\rho(k_\tau q_{+f\tau}k_\tau^{-1},k_\tau q_{-f\tau}k_\tau^{-1})=\rho(q_{+f\tau},q_{-f\tau})$ for each pair $(f,\tau)$ and $k_\tau\in\SU(2)$. The previous class of models is recovered by choosing a coupling of the special form: $\rho(q_+,q_-) = \delta(q_+ q_-^{-1}) \mu(q_+)$ for the geometric sector. Notice however that one can now switch between the geometric and non-geometric sectors only with a change of measure $\tl{\rho}(q_+,q_-) = \rho(q_+,q_-^{-1})$. This is indeed equivalent to replacing $q_{-f\tau}$ in the action with it inverse. Thus, the boundary between the two sectors is weakened by the weakening of the constraints.

Since $\rho$ can be expanded into matrix elements of $q_+$ and $q_-$ contracted with intertwiners, let us restrict attention to $\rho(q_{+f\tau},q_{-f\tau}) = D^{(j_{+f,\tau})}_{A_{+f,\tau} B_{+f\tau}}(q_{+f\tau}) D^{(j_{-f,\tau})}_{A_{-f,\tau} B_{-f\tau}}(q_{-f\tau})$. One can then perform the integrals over $q_{\pm f\tau}$ using the orthogonality relation \eqref{orthogonality}, which enforces the equalities $j_{+f\tau} = j_{+f}$ and $j_{-f\tau} = j_{-f}$. Upon absorbing the rotations $h_\tau$ into the holonomies $g_{+v\tau}$ as in the previous models, we have:
\be \label{weak general}
Z = \int \prod_{(\tau,v)} dG_{v\tau}\, \prod_f d_{j_{+f}} d_{j_{-f}}\, \prod_\tau \f{1}{\prod_{f\subset \pp\tau} d_{j_{+f}} d_{j_{-f}}}\, \prod_v \prod_{f\subset \pp v} D^{(j_{+f})}_{A_{+f,\tau} B_{+f,\tau'}} \Bigl(g_{+v\tau}^{-1} g_{+v\tau'}\Bigr)\,D^{(j_{-f})}_{A_{-f,\tau} B_{-f,\tau'}} \Bigl(g_{-v\tau}^{-1} g_{-v\tau'}\Bigr)
\ee
where $\tau$ and $\tau'$ are the two tetrahedra sharing the triangle $f$ in the 4-simplex $v$. It is thus clear that the function $\rho$ directly encodes the couplings of the model. The features which do not depend on the choice of $\rho$ are the labelling of faces with Spin(4) representations $(j_{+f},j_{-f})$, the face amplitude $d_{j_{+f}} d_{j_{-f}}$, and the fact that the vertex is a product over the triangles of $f$ of functions of $g_{+v\tau}^{-1} g_{+v\tau'}$ and $g_{-v\tau}^{-1} g_{-v\tau'}$.

Let us now show that \eqref{weak general} is a general setting allowing to understand the new vertices as weakly imposing the constraints. Indeed, we are already assured to be able to write the vertex with the fusion coefficients. Moreover, in \eqref{weak general}, the representations labelling faces are now exactly those coming from the expansion of $\rho$. We thus look for the above-mentioned coefficients $C^{j_f}_{j_f}{}^{\phantom{-}j_f}_{-j_f}{}^{k_{f\tau}}_0$. Consider the special choice:
\be \label{epr measure}
\rho (q_+,q_-) = \sum_j \int d^2\hat{n}\ \bra j,\hat{n}\rvert q_+ \lvert j,\hat{n}\ket \bra j,\hat{n}\rvert q_- \lvert j,\hat{n}\ket
\ee
where $\lvert j,\hat{n}\ket$ are $\SU(2)$ coherent states, here associated to pairs $(f,\tau)$. They have been introduced in \cite{coherent states} and further used because their semi-classical behaviour eases the geometrical interpretation of the constraints. However, they do not seem to naturally appear in our computations. The function \eqref{epr measure}, which leads to the flipped model of \cite{EPR} (or the unflipped model by choosing $\rho(q_+,q_-^{-1})$), can be compared with that corresponding to the BC model:
\be \label{BC measure}
\rho_{BC} (q_+,q_-) = \delta(q_+ q_-^{-1}) = \sum_j d_j^2 \int d^2\hat{n}\ \bra j,\hat{n}\rvert q_+\ q_-^{-1} \lvert j,\hat{n}\ket
\ee
which shows that \eqref{epr measure} indeed weakens the natural $\SU(2)$ measure corresponding to the BC model.

Expanding \eqref{epr measure} into matrix elements, we obtain the following expression:
\begin{gather}
Z = \sum_{\{j_f\}} \int \prod_{(f,\tau)} d^2\hat{n}_{f,\tau}\ \prod_f d_{j_f}^2\ \prod_\tau \f{1}{\prod_{f\subset\pp\tau} d_{j_f}^2}\ \prod_v A_v \\
\text{with}\quad A_v = \int \prod_{\tau\subset\pp v} dG_{v\tau}\ \prod_f \bra j_f,\hat{n}_{f\tau'}\rvert g_{+\tau'v}\ g_{+v\tau}\lvert j_f,\hat{n}_{f\tau}\ket\ \bra j_f,\hat{n}_{f\tau'}\rvert g_{-\tau'v}\ g_{-v\tau}\lvert j_f,\hat{n}_{f\tau}\ket
\end{gather}
This expression has been shown in \cite{FKLS} to reproduce the flipped model of \cite{EPR}. The other sector, that is the unflipped model of \cite{consistently solving, FKLS}, can be obtained with a simple change of measure which leads to:
\be
A_v = \int \prod_{\tau\subset\pp v} dG_{v\tau}\ \prod_f \bra j_f,\hat{n}_{f\tau'}\rvert g_{+\tau'v}\ g_{+v\tau}\lvert j_f,\hat{n}_{f\tau}\ket\ \bra j_f,\hat{n}_{f\tau}\rvert g_{-v\tau}^{-1}\ g_{-\tau'v}^{-1}\lvert j_f,\hat{n}_{f\tau'}\ket
\ee

An important part of the content of these models has been here introduced by the measure $\rho$, used to weaken the constraints. The other physical inputs, that is the choice of the variables, and the imposition of the constraints on disjoint tetrahedra, are common features shared with the BC model. From this point of view, a better understanding may require a precise analysis of the function $\rho$ in \eqref{epr measure}. Using $\bra j,\hat{n}\rvert g\lvert j,\hat{n}\ket = (\cos\theta + i\sin\theta\ \hat{u}\cdot\hat{n})^{2j}$ for $g = \cos\theta + i\sin\theta\ \hat{u}\cdot\vec{\sigma}$, the integral over the unit vector $\hat{n}$ can be explicitly performed with repeated use of the binomial formula. $\rho(q_+,q_-)$ then appears as a function of the class angles of the $\SU(2)$ group elements $q_+$ and $q_-$, and of the dot product of their directions. The resulting expression seems however quite cumbersome.

\section{Physics of the BC model and beyond} \label{beyond BC}

It is the first time, as far as we know, that the BC model is derived using a path integral formulation on a lattice. We thus focus in this section on the physical setting of the model. The variables have clearly identified geometric roles. The clean treatment of the situation points out the drawbacks of the BC model with regards to quantum gravity, which can be traced back to the issue of gluing between tetrahedra, and thus to that of their correlations.

The BC model \cite{BC} has been originated by considering an isolated 4-simplex, and imposing the constraints on each of its tetrahedra. It has then been thought that the failure of the BC model to describe quantum gravity resides in this specific fact, and that imposing the constraints on a complete triangulation leads to the new vertices, depending on the sector, \cite{FKLS}. We have however seen here that the BC model can be thought of on a triangulation (discarding boundary terms), without isolating 4-simplices from each other, and that the new spin foam models can also be considered in this framework. A rotation $h_\tau$ is associated to each tetrahedron and geometrically represents the direction $N_\tau=(h_\tau,\mathrm{id})\triangleright N^{(0)}$ perpendicular to $\tau$, without refering to the 4-simplices sharing $\tau$. The fact that the BC model deals with manifestly decorrelated vectors $N_{\tau,v}$ arises from the change of variables \eqref{decoupling normals}. The rotations $h_\tau$ are defined at each tetrahedron, but the amplitude, comparing the ways two adjacent tetrahedra map $N^{(0)}$ into their normal directions, takes place at each 4-simplex. Thus, integrating over the holonomy degrees of freedom decorrelates $h_\tau(s(\tau))$ and $h_\tau(t(\tau))$. This is in contrast with the derivation of the BC model proposed in \cite{BC, FKLS}, in which the constraints are imposed 4-simplex by 4-simplex (and then tetrahedron by tetrahedron). Moreover, the form of the simplicity constraints used in \cite{FKLS} to derive the new vertex is the same as that which has been here shown to reproduce the BC model when explicitly integrating bivectors.

It is clear by looking at the equations of motion of the BC action that the BC model contains some correlations between neighbouring tetrahedra $\tau_1$ and $\tau_2$. Such correlations are given by \eqref{corr BC} on-shell, while the kernel of the BC model \eqref{BC} is due to comparisons between the rotations $h_{\tau_1}$ and $h_{\tau_2}$. However, we have seen in section \ref{sec BC} that the latter are directly related by the simplicity constraints. Indeed, the cross-simplicity constraints are solved by assigning rotations $h_\tau$ to tetrahedra, but without forgetting that for tetrahedra sharing a triangle, these normals are correlated since they live on the plane orthogonal to $\star B_f$. Therefore, correlations between tetrahedra are intimately related to the gluing process. These correlations are not correctly taken into account in the BC model, because of the gluing and of the fact that the rotations $h_\tau$ are considered as independent variables.

The correlations are written in \eqref{coupling tetra} in terms of elements $\phi_{f,\tau}$ living in the $\U(1)$ subgroup preserving the self-dual part $b_{+f}$ of each bivector. The $\U(1)$ subgroup preserving $b_{+f}$ is of major physical importance because it enables to relate the normal directions of neighbouring tetrahedra, and thus describes the whole set of orthogonal vectors to a given triangle. Let us introduce bivectors $B_{f\tau}$ like in section \ref{sec BC}. Then we can choose a normal $N_{f,\tau}$ of reference for each tetrahedron sharing $f$, encoded in $h_{f\tau}\in\SU(2)$, to impose diagonal simplicity for each piece $(f,\tau)$: $b_{-f\tau} = -h_{f\tau}^{-1}\,b_{+f\tau}\,h_{f\tau}$. Cross-simplicity is very easy to impose in this framework:
\be
h_{f\tau} = e^{i\alpha_{f\tau}\,\hat{b}_{+f\tau}\cdot\f{\vec{\sigma}}{2}}\ h_\tau
\ee
with $\alpha_{f\tau}\in[0,4\pi)$. $\hat{b}_{+f\tau}$ is the direction of $b_{+f\tau}$, so that $h_{f\tau}h_\tau^{-1}$ preserves $b_{+f\tau}$. We then need to introduce some explicit rules for parallelly transporting $B_{f\tau}$ between adjacent tetrahedra. The norm $\lv b_{+f\tau}\rv = \lv b_{-f,\tau}\rv$ is the same for all $\tau$, and $G_{\tau \tau'}$ sends the directions $(\hat{b}_{+f\tau},\hat{b}_{-f\tau})$ of $B_{f\tau}$ to those of $B_{f\tau'}$. One can show that, as far as only the directions are concerned, it is equivalent to give such rules for the self-dual parts $b_{+f\tau}$ together with a rule to relate the vectors $N_{f\tau}$ between tetrahedra sharing $f$. Not surprisingly, the latter is of the form \eqref{coupling tetra}:
\be
h_{f\tau}\,h_{f\tau'}^{-1}(\tau) = h_{f\tau}\,g_{-\tau\tau'}\,h_{f\tau'}^{-1}\,g_{+\tau\tau'}^{-1} = e^{i\psi^f_{\tau\tau'}\,\hat{b}_{+f\tau}\cdot\f{\vec{\sigma}}{2}}
\ee
with $\psi^f_{\tau\tau'}\in[0,4\pi)$. Again, the r.h.s. leaves $b_{+f\tau}$ invariant. It means that the issue of parallelly transporting $B_{f\tau}$ is intimately related to that of consistently imposing simplicity.

Another problem of the BC model, as already mentioned in the study of the equations of motion of the BC action, is that it does not reproduce any discrete analog of Einstein's equations. We expect in particular the following relation: $G_f(\tau) B_{f\tau} G_f^{-1}(\tau) = B_{f\tau}$, where $G_f$ is the holonomy around the dual face $f$, as a discretization of the fact that $(e^I\wedge e^J)\wedge (\star F)_{JK} =0$ in the continuum (or from the e.o.m. $d_AB^{IJ}=0$ as already discussed). This relation between the triangles and the holonomies around them can be seen in terms of the vectors $N_{f\tau}$. Indeed, $G_f(\tau)\in\U(1)_{b_{+f\tau}}\times \U(1)_{b_{-f\tau}}$ implies that though $G_f(\tau)$ does not leave $N_{f\tau}$ invariant as in \eqref{trivial eom}, it sends it instead onto a new vector which still lies on the plane orthogonal to $\star B_f(\tau)$:
\be
\eps_{IJKL}\ \Bigl[G_f(\tau)\triangleright N_{f\tau}\Bigr]^J\ B_{f\tau}^{KL} = 0
\ee
The analog of Einstein's equations in the Plebanski formulation is obtained by extremizing the action w.r.t. the field $B$. In the BC action, this gives equations \eqref{trivial eom} which are independent of bivectors and require that the normal $N_\tau$ is left invariant by $G_{f\tau}$. This is due to the fact that the action is built on disjoint tetrahedra and is linear in the bivectors. However, the above discussion shows that the directions and the norms of $b_{\pm f\tau}$ have quite different roles. A study of a discrete action built on these ideas and of the resulting spin foam model will appear in a coming paper \cite{coming}.

\section*{Conclusion}

We have shown that the Barrett-Crane spinfoam model can be derived from a discrete action
principle and is related to a quantization of the geometric (gravitational) sector of Plebanski's theory. It is based as usual on a BF term discretized on a triangulation.
The simplicity constraints are independently imposed on each tetrahedron and we explicitly perform the integrals over the bivectors. The Barrett-Crane spin foam amplitude is recovered for a specific choice of measure weakly imposing diagonal and cross-simplicity. This measure can be translated into the use of a non-commutative composition of plane waves, and it is argued to be the natural way to proceed in spin foam computations. We have also proposed an action which sticks tetrahedra together using the $\SU(2)$ group structure. This method straightforwardly leads to the same results and thus implicitly takes into account the above-mentioned measure.

It is known that the BC model does not precisely correspond to gravity since it does not reproduce the expected correlations in semi-classical computations of graviton type \cite{BC fail}. The Engle-Pereira-Rovelli (EPR) and Freidel-Krasnov (FK) models have been proposed to remedy the drawbacks of the BC model. We have here obtained a Lagrangian formulation of these models by weakening the implementation of the constraints in the lattice path integral. In our view a better understanding of these models requires a careful study of the regluing process, which sticks tetrahedra together after imposition of the constraints. The regluing which enables to recover the BC model is that usually considered in pure BF theory and {\it unconstrained} BF-like theories \cite{BF action principle}. However, the Lagrange multipliers can be seen as a source of curvature, so that this regluing yield incorrect rules for parallel transport of bivectors. This issue is related to the geometric consistency of the constraints and to a faithful accounting for correlations between neighbouring tetrahedra. These correlations are detailed in section \ref{beyond BC}. Equation \eqref{coupling tetra}, whose trace gives the dot product of the normals to adjacent tetrahedra and thus contains the dihedral angles of Regge calculus, describes them in terms of variables living in $\U(1)\times\U(1)$ subgroups of Spin(4) which leave each
bivector invariant. A proposition using precisely such variables is
under preparation \cite{coming}.

In the non-geometric sector, the amplitude is built on the 15j-symbol for simple unitary irreducible representations of Spin(4), exactly like the model only implementing diagonal simplicity. However, they both differ due to the amplitudes associated to triangles and tetrahedra. A precise introduction of the Immirzi parameter into the partition function is also proposed and shown to differ from its usual introduction in a measure factor.

Finally, the fusion coefficients, at the root of the EPR and FK spin foam models, are derived starting from the setting of the BC model supplemented with a non-trivial measure for the Lagrange multipliers. Thus, they appear to be very natural objects in the context of spin foam quantization of gravity. From this point of view, the main questions concern the spin data labelling the links of the boundary spin networks and the summation of these 4-simplex amplitudes, that is to say the choice of weights for triangles and tetrahedra.

\section*{Acknowledgements}

The authors are grateful to Mait\'e Dupuis for her active participation in the early stages of this work.

\appendix
\section{Some useful formulas} \label{formula}

$\SU(2)$ group elements are parametrized by : $g = \cos\theta + i\sin\theta\ \hat{n}\cdot\vec{\sigma}$, where $\sigma$ are the Pauli matrices satisfying $[\sigma_i,\sigma_j]=2i\eps_{ij}^{\phantom{ij}k}\sigma_k$. We define the projection of $g$ onto the Pauli matrices $\vec{p} = \sin\theta\ \hat{n}$. Thus, the (unnormalized) Haar measure over $\SO(3)$ (restricting the parametrisation to $\theta\in [0,\f{\pi}{2}]$) is related to the Lebesgue measure on $\mathbbm{R}^3$ :
\be
dg_{\SO(3)} = \f{d^3p}{\sqrt{1-\vec{p}^2}}
\ee
The factor $(1-\vec{p}^2)^{\f{1}{2}}$ restores the compactness of the group when working with the $\vec{p}$ variables, $\lvert\vec{p}\rvert<1$.

Using test-functions, it is easy to show that : $\delta_{\SO(3)}\big(g h^{-1}\big) = \sqrt{1-\vec{p}_g^2}\ \delta^{(3)}(\vec{p}_g-\vec{p}_h)$. Rewriting the delta over $\mathbb{R}^3$ with the usual Fourier transform, we have :
\be \label{int b}
\int d^3b\ e^{i\big[\mathrm{tr}(bg)-\mathrm{tr}(bh)\big]} = \f{1}{\sqrt{1-\vec{p}_g^2}}\Big( \delta_{\SU(2)}\big(g h^{-1}\big)+\delta_{\SU(2)}\big(-g h^{-1}\big)\Big) = \f{1}{\lvert\tr\ g\rvert}\ \delta_{\SO(3)}\big(g h^{-1}\big)
\ee
where $b=b^i\tau_i$ is a Lie algebra $\su(2)$ element, integrated over with the Lebesgue measure, with $\tau_i=-\f{i}{2} \sigma_i$. Such a formula does not admit a straightforward generalisation as soon as more than two group elements are involved. Indeed, a sum of $\vec{p}$ variables, $\vec{p}_{h_1}+\vec{p}_{h_2}$, is not in general the projection of a group element onto the Pauli matrices. Equation \eqref{int b} is very simple, but given the parametrisation of the constraints introduced in section \ref{sec simple rep}, it is the only formula needed in this work to explicitly perform the integrals over the bivectors.

After the integrals over the bivectors, one encounters integrals of products of matrix elements of $\SU(2)$ representations. The basic formulas are that expressing the orthogonality of these matrix elements,
\be \label{orthogonality}
\int dg\ D^{(j_1)}_{ab}(g)\ D^{(j_2)*}_{cd}(g) = \f{1}{d_{j_1}}\delta_{j_1j_2}\ \delta_{ac}\delta_{bd},
\ee
and that giving the decomposition of a tensor product of representations,
\be \label{tensor prod}
D^{(j_1)}_{ab}(g)\ D^{(j_2)}_{\alpha\beta}(g) = \sum_{J,A,B} C_a^{j_1}{}_\alpha^{j_2}{}_A^J\ C_b^{j_1}{}_\beta^{j_2}{}_B^J\ D^{(J)}_{AB}(g).
\ee
with $\lvert j_1-j_2\rvert\leq J\leq j_1+j_2$ and $-J\leq j_1,j_2\leq J$, and where the coefficients $C^{j_1}_{.}{}^{j_2}_{.}{}^{J}_{.}$ are Clebsch-Gordan coefficients between the irreducible representations $j_1\otimes j_2 \rightarrow J$. One can easily show, using \eqref{orthogonality} and \eqref{tensor prod}, that:
\begin{gather} \label{int g3}
\int dg\ D^{(j_1)}_{ab}(g)\ D^{(j_2)}_{cd}(g)\ D^{(j_3)*}_{ef}(g) = \f{1}{d_{j_3}} C_a^{j_1}{}_c^{j_2}{}_e^{j_3}\ C_b^{j_1}{}_d^{j_2}{}_f^{j_3} \\
\int dg\ D^{(j_1)}_{ab}(g)\ D^{(j_2)}_{cd}(g)\ D^{(j_3)*}_{\alpha\beta}(g)\ D^{(j_4)*}_{\gamma\delta}(g) = \sum_{J,A,B} \f{1}{d_J} C_a^{j_1}{}_c^{j_2}{}_A^{J}\ C_b^{j_1}{}_d^{j_2}{}_B^{J}\ C_\alpha^{j_3}{}_\gamma^{j_4}{}_A^{J}\ C_\beta^{j_3}{}_\delta^{j_4}{}_B^{J}
\end{gather}
One can express these relations with the help of the 3jm-symbols which enjoy better symmetries than the Clebsch-Gordan coefficients thanks to:
\be
C^{j_1}_{m_1}{}^{j_2}_{m_2}{}^{j_3}_{m_3} = (-1)^{m_3+j_1-j_2} \sqrt{d_{j_3}}\ \begin{pmatrix} j_1 & j_2 & j_3 \\ m_1 & m_2 & -m_3 \end{pmatrix}
\ee

\section{Another derivation of the BC model} \label{naive BC}

We show how cross-simplicity can be imposed on disjoint tetrahedra in the most naive way, that is by simply adding them to the action. In this framework, the BC model corresponds to a specifically weak imposition of the contraints corresponding to the measure \eqref{bessel measure}, and not \eqref{broken up measure}. This feature is naturally hidden when we stick the BF action with the constraints using the $\SU(2)$ multiplication, as in section \ref{sec BC}, a process which respects the $\SU(2)$ structure of the spin foam.

To impose the constraints independently on each tetrahedron, the triangulation has been first broken up into a disjoint union of tetrahedra. Then, we associate bivectors $B_{f\tau}$ to each face of each tetrahedron, and consider them to be all independent from each other. We will then impose the cross-simplicity constraints by asking for the equality of the rotations associated to the triangles of a tetrahedron.

Let us first build the BF action using the variables $B_{f\tau}$. The construction is similar to that using the wedges. From the point of view of the 2-complex dual to the triangulation, a wedge is a pair $(f,v)$ formed by a face and one of its vertices (i.e. a triangle seen in a particular 4-simplex, in the triangulation language). The variables $(f,\tau)$ that we use here are pairs consisting of a face and one of its boundary edges (i.e. a triangle seen in a particular tetrahedron): draw segments from the center of each dual face towards each of its vertices, as shown figure \ref{variables Bftau}. We thus use decorrelated bivectors $B_{f\tau}$ for each pair $(f,\tau)$. The BF action then couples them to some 'holonomies' $G_{f\tau}$ around the pieces $(f,\tau)$. Two pieces $(f,\tau_1)$ and $(f,\tau_2)$, for tetrahedra $\tau_1$ and $\tau_2$ sharing $f$ in the same 4-simplex $v$, have a common link delimiting them (see figure \ref{variables Bftau}), and which carries a Spin(4) element, $L_{f,v}$ entering the elements $G_{f\tau_{1,2}}$. Integrating these variables in the path integral can be seen as a regluing process. As discussed in section \ref{eom BC}, this regluing perfectly works as far as there are no sources of curvature, which is the case in BF theory but not in our action for the BC model. The following action reproduces the usual spin foam model for BF theory:
\be
S_{BF} = \sum_{(f,\tau)} \Tr \big( B_{f\tau}\ G_{f\tau}\big)
\ee
As noticed in section \ref{sec BC}, this action is {\it a priori} different from the standard discrete BF action \eqref{BF action}. The particularity of BF theory that it is stable under breaking up the triangulation into tetrahedra, 4-simplices or both is directly due to the fact that it deals with trivial parallel transport around dual faces.

Let us now present the computations of the partition function from the action $S_{BC}$, in which the constraints for the geometric sector are naively introduced:
\be
S_{BC} = \sum_{(f,\tau)} \Tr \big( B_{f\tau}\ G_{f\tau}\big) + \Tr \big[q_{f\tau} (b_{-f\tau}+ h_\tau^{-1}\, b_{+f\tau}\, h_\tau)\big]
\ee
where $B_{f\tau}$, $q_{f,\tau}$ and $h_\tau$ are defined in the local frame of $\tau$. $S_{BC}$ is a function of the Spin(4) group elements $G_{v\tau}$ (collectively denoting the basic variables $G_{s(\tau)\tau}$ and $G_{\tau t(\tau)}$) and $L_{fv}$, the $\SU(2)$ Lagrange multipliers $q_{f\tau}$, the rotations $h_\tau$ standing for the normals $N_\tau$ to each tetrahedron, and of the bivectors $B_{f\tau}$. The partition function thus reads:
\begin{gather}
Z_{\mu} = \int \prod_{(v,\tau)} dG_{v\tau} \prod_{(f,v)} dL_{fv} \int \prod_\tau dh_\tau \int \prod_{(f,\tau)} dB_{f\tau}\ \ e^{iS_{BF}}\ \prod_{(f,\tau)} \tl{\delta}_\mu\bigl(b_{-f\tau}+ h_\tau^{-1}\, b_{+f\tau}\, h_\tau\bigr) \\
\text{with}\qquad \tl{\delta}_\mu(x_{f\tau}) = \int_{\SU(2)} dq_{f\tau}\ \mu(q_{f\tau})\ \exp\big\{i\,\tr[q_{f\tau}\,x_{f\tau}]\big\}
\end{gather}

As tetrahedra have been unstuck, integrating over the bivectors is very easy with \eqref{int b} and makes the measure factors $\lvert \tr(g)\rvert$ appear. Then, the integrals over the Lagrange multipliers relate the self-dual parts of the elements $G_{f\tau}=(g_{+f\tau},g_{-f\tau})$ to their anti-self-dual parts:
\be \label{intermediate step}
Z_{\mu} = \int \prod_{(\tau,v)} dG_{v\tau} \prod_{(f,v)} dL_{fv} \int \prod_\tau dh_\tau\
\prod_{(f,\tau)}\f{\mu(g_{-f\tau})}{(\tr\,g_{-f\tau})^2}\ \delta\Bigl(g_{+f\tau}\ h_\tau\
g_{-f\tau}^{-1}\ h_\tau^{-1}\Bigr)
\ee
Note that, similarly to \eqref{physical step simple rep}, the group element $g_{+f\tau}\, h_\tau\, g_{-f\tau}^{-1}$ is a kind of parallelly transported $G_{f\tau}\triangleright h_\tau$ of the rotation $h_\tau$, so that $G_{f\tau}$ has to preserve the vector $N_\tau$. This restriction on the fields is responsible for the non-topological character of the model, and is independent of the measure $\mu$ of the Lagrange multpliers. It could have been in fact anticipated from the action \eqref{BC action} since the latter is linear in the bivectors, so that equation \eqref{intermediate step} simply expresses the projection onto the corresponding equations of motion.

We now restrict attention to the special choice \eqref{bessel measure} for the measure $\mu$. This amounts to considering only the delta functions over the group in \eqref{intermediate step}, and equivalently to use the specific function $\tl{\delta}_{\mu_{BC}}(\vec{x}) = \f{J_1(\lvert\vec{x}\rvert)+ J_3(\lvert\vec{x}\rvert)}{\lvert\vec{x}\rvert}$ for each pair $(f,\tau)$, instead of the strong condition $\delta^{(3)}(\vec{x})$. Integrating the boundary connection variables $L_{fv}$ is then easily done since each of them only appears twice. In particular it implies the equality of the representations $j_{f\tau}$, assigned to the pairs $(f,\tau)$ via the expansion of the delta functions into characters, to a single representation $j_f$ for each face. A triangle in a single 4-simplex $v$ being shared by exactly two tetrahedra, let us denote them $u(v)$ and $d(v)$. The partition function is thus:
\be \label{BC int step 2}
Z_{BC} = \int \prod_{(\tau,v)} dG_{\tau v} \int \prod_\tau dh_\tau \sum_{\{j_f\}} \prod_f d_{j_f}^2 \prod_\tau \f{1}{\prod_{f\subset\pp\tau}d_{j_f}}\ \prod_v \Bigl[ \prod_{f\subset \pp v} \chi_{j_f}\bigl(h_{u(v)}(v)\ h^{-1}_{d(v),v}(v)\bigr)\Bigr]
\ee
The rotation $h_\tau(v)$ stands for the parallelly transported of $h_\tau$ to the 4-simplex $v$ along the boundary of a dual face. In \eqref{BC int step 2}, $h_\tau$ is only transported to its source vertex (in the dual picture) $s(\tau)$ and its target vertex $t(\tau)$: $h_\tau(s(\tau)) = g_{+s(\tau)\tau}\ h_\tau\ g_{-s(\tau)\tau}^{-1}$ and $h_\tau(t(\tau)) = g_{+\tau t(\tau)}^{-1}\ h_\tau\ g_{-\tau t(\tau)}$. It means that the vector $N_\tau$, orthogonal to the tetrahedron $\tau$ and defined in its rest-frame, is parallelly transported to the two frames of the 4-simplices sharing it. Then the character $\chi_{j_f}(h_{u(v)}(v) h^{-1}_{d(v)}(v))$ compares the way the vector $N^{(0)}$ is mapped into $N_\tau$ for two adjacent tetrahedra at the 4-simplex $v$. Since the self-dual components $g_{+v\tau}$ of the group elements realizing these transports only appear here in the amplitude, $g_{-v\tau}$ and $h_\tau$ can be absorbed into them by the change of variables \eqref{decoupling normals}. This directly leads to the integral representation of the 10j-symbol, \eqref{BC}.

\section{The BC model in the non-geometric sector} \label{app topo sector}

The model turns out to be ill-defined if we simply think of imposing $b_{-f\tau}-h_\tau^{-1} b_{+f\tau} h_\tau=0$ by changing $q_{f\tau}$ with its inverse in the self-dual part of the action \eqref{BC star action}. It indeed crucially depends on the parity of the number of dual edges for each dual face. The model becomes well defined if we use the wedge variables, superimposed to the variables $(f,\tau)$ previously used, according to figure \ref{half-wedges}. Let us stress that all the models derived in this work can be reformulated using these half-wedge variables. However, since they are only necessary in the present case, we have chosen not to use them before to simplify the constructions and to obtain more straitghforward interpretations.

Let us break up the triangulation into a disjoint union of 4-simplices. Thus, there exist as many copies of a given triangle as it is shared by different 4-simplices. In the dual picture, it is represented by a division of each dual face into pieces corresponding to the choice of a boundary vertex, that is into pieces $(f,v)$ called wedges. Let us now break up each 4-simplex into tetrahedra. Since a tetrahedron is shared by two 4-simplices, it means that the number of copies of a triangle is twice that obtained with the wedges. Thus, with regards to the above-mentioned parity problem of the number of dual edges, this construction will select the case where each face has an even number of edges, since all happens as if tetrahedra were chopped into two pieces. A copy of a face is specified by the data of a tetrahedron and a 4-simplex containing it. The corresponding division of a dual face is presented in figure \ref{half-wedges}, and we call each piece $(f,\tau,v)$ a half-wedge.

We are now looking for imposing the constraints on each half-wedge, but with the same rotation $h_\tau$ for half-wedges of the same face and same tetrahedron, that is imposing the existence of a unique normal $N_\tau$ such that $N_{\tau,I}B_{f\tau v}^{IJ}=0$ for the two copies of $\tau$. In the naive way to impose the constraints on the action, bivectors have to be independently assigned to each copy of a triangle, $B_{f\tau v}$, and the regluing of the simplices can be partially done with boundary variables associated to the links delimiting the pieces $(f,\tau,v)$ (the dashed and dotted lines in figure \ref{half-wedges}). But this can be done more efficiently with the arguments leading to \eqref{BC star action}. Having multipliers $q_{f\tau v}$ to impose the constraints on each half-wedge, these can be viewed as sources $Q_{f\tau v} = (h_\tau q_{f\tau v}h_\tau^{-1},q_{f\tau v})$ inserted into the holonomies $G_f$ for the geometric sector. The flipping of the sign from \eqref{diag sol} to \eqref{topo sector simple rep} translates into the change $q_{f\tau v}\rightarrow q_{f\tau v}^{-1}$ in the self-dual part of the action. We thus define $\tl{Q}_{f\tau v} = (h_\tau q^{-1}_{f\tau v}h_\tau^{-1},q_{f\tau v})$. Obviously, for the geometric sector, one can define the variables $q_{f\tau}\equiv q_{f\tau s(\tau)} q_{f\tau t(\tau)}$ and $Q_{f\tau}\equiv Q_{f\tau s(\tau)} Q_{f \tau t(\tau)}$ appearing in the action of the BC model \eqref{BC star action}. This is however impossible in the non-geometric sector, because of the flipping $q_{f \tau v}\rightarrow q_{f\tau v}^{-1}$ in the self-dual part of the action.

Let us mention a subtlety of the construction. Since a tetrahedron exists in two copies in the shattered triangulation, these copies have {\it a priori} different frames: one for $\tau$ in the 4-simplex $s(\tau)$ and one for $\tau$ in $t(\tau)$. We thus introduce a Spin(4) element $G_\tau^{s(\tau)t(\tau)}$ responsible for the parallel transport between these two frames. The rotations describing the same normal in the two frames are related by: $h_\tau^{(s(\tau))} = g_{+\tau}^{s(\tau)t(\tau)} h_\tau^{(t(\tau))} g_{-\tau}^{s(\tau)t(\tau)-1}$. This relation supplements the following action:
\be \begin{split}
\tl{S}_{C-S} = \sum_f &\tr\Big(b_{-f}\ g_{-v_0 \tau_1}\ q_{f,\tau_1,v_0}\ g_{-\tau_1}^{v_0 v_1}\ q_{f,\tau_1,v_1}\ g_{-\tau_1 v_1}\ g_{-v_1 \tau_2}\ q_{f,\tau_2,v_1}\cdots\Big) \\
 &+\tr\Big(b_{+f}\ g_{+v_0 \tau_1}\ h_{\tau_1}^{(v_0)} q_{f,\tau_1,v_0}^{-1}\ g_{-\tau_1}^{v_0 v_1}\ q_{f,\tau_1,v_1}^{-1} h_{\tau_1}^{(v_1)\ -1}\ g_{+\tau_1 v_1}\ g_{+v_1 \tau_2}\ h_{\tau_2}^{(v_1)} q_{f,\tau_2,v_1}^{-1}\cdots\Big)
\end{split} \ee
This action being linear in the bivectors, integrating them in the partition function is easy:
\be \begin{split}
\tl{Z} = \int \prod_\tau dG_{s(\tau)\tau} dG_{\tau t(\tau)} dG_\tau^{s(\tau) t(\tau)} \int \prod_{(f,\tau,v)} &dq_{f,\tau,v} \prod_{(\tau,v)} dh_\tau^{(v)}\ \prod_f \delta\Big(g_{-v_0 \tau_1}\ q_{f,\tau_1,v_0}\ g_{-\tau_1}^{v_0 v_1}\ q_{f,\tau_1,v_1}\ g_{-\tau_1 v_1}\ g_{-v_1 \tau_2}\ q_{f,\tau_2,v_1}\cdots\Big) \\
&\delta\Big(g_{+v_0 \tau_1}\ h_{\tau_1}^{(v_0)} q_{f,\tau_1,v_0}^{-1}\ g_{-\tau_1}^{v_0 v_1}\ q_{f,\tau_1,v_1}^{-1} h_{\tau_1}^{(v_1)\ -1}\ g_{+\tau_1 v_1}\ g_{+v_1 \tau_2}\ h_{\tau_2}^{(v_1)} q_{f,\tau_2,v_1}^{-1}\cdots\Big)
\end{split} \ee

Expanding the delta functions introduces representations $j_{+f}$ and $j_{-f}$ labelling faces. Integrating the $\SU(2)$ variables $q_{f\tau v}$ is very standard, using the orthogonality relation \eqref{orthogonality}. It enforces the equalities $j_{+f}=j_{-f}=j_f$, eliminates the rotations $h_\tau^{(v)}$, and recombines the holonomies around the dual faces $G_f = G_{v_0\tau_1}G_{\tau_1}^{v_0 v_1}G_{\tau_1 v_1}\cdots$, exactly as explained in section \ref{non-geom sector}. We can thus turn to the variables $G_\tau\equiv G_{s(\tau)}G_\tau^{s(\tau)t(\tau)}G_{\tau t(\tau)}$ representing parallel transport between the 4-simplices $s(\tau)$ and $t(\tau)$. It means in particular that the elements $G_\tau^{s(\tau)t(\tau)}$ can be gauge fixed to the identity, so that the action can be simplified to that presented in section \ref{sec BC}, \eqref{topo action}. One finally ends up with the spin foam model of equation \eqref{topo model}, whose difference with the model for BF theory restricted to Spin(4) simple representations, implementing simplicity of each bivector independently and presented in section \ref{sec simple rep}, lies in the face and tetrahedron amplitudes.


\end{document}